\documentstyle[tighten,preprint,eqsecnum,aps,prd]{revtex}
\begin{document}
\draft

\hyphenation{
mani-fold
mani-folds
}


\def\Bbb{\bf}

\def\BbbR{{\Bbb R}}
\def\BbbZ{{\Bbb Z}}

\def\casehalf{{\case{1}{2}}}

\def\ads{{anti-de~Sitter}}
\def\rnads{{Reissner-Nordstr\"om-anti-de~Sitter}}

\def\bm{{\bf m}}
\def\bq{{\bf q}}

\def\Rh{R_{\rm h}}
\def\Rhor{R_{\rm hor}}
\def\Rcrit{R_{\rm crit}}

\def\czren{{\cal Z}_{\rm ren}}
\def\zren{Z_{\rm ren}}


\preprint{\vbox{\baselineskip=12pt
\rightline{PP96--63}
\rightline{WISC--MILW--96--TH--11}
\rightline{gr-qc/9602003}}}
\title{Hamiltonian thermodynamics of the
Reissner-Nordstr\"om-anti-de~Sitter
black hole\footnote{Dedicated to Karel Kucha\v{r}
on the occasion of his sixtieth birthday.
({\it Physical Review\/} D did not, alas,
permit a dedication in the published version of this paper.)
}}
\author{Jorma Louko\footnote{On leave of absence from
Department of Physics, University of Helsinki.
Electronic address:
louko@wam.umd.edu}}
\address{
Department of Physics,
University of
Wisconsin--Milwaukee,
\\
P.O.\ Box 413,
Milwaukee, Wisconsin 53201, USA
\\
and
\\
Department of Physics,
University of Maryland,
College Park,
Maryland 20742--4111,
USA\footnote{Present address.}}
\author{Stephen N. Winters-Hilt\footnote{Electronic address:
winters@csd.uwm.edu}}
\address{
Department of Physics,
University of
Wisconsin--Milwaukee,
\\
P.O.\ Box 413,
Milwaukee, Wisconsin 53201, USA}
\date{Revised version, June 1996. Published in
{\it Phys.\ Rev.\ \rm D \bf 54} (1996), pp.~2647--2663.}
\maketitle
\begin{abstract}%
We consider the Hamiltonian dynamics and thermodynamics of spherically
symmetric Einstein-Maxwell spacetimes with a negative cosmological
constant. We impose boundary conditions that enforce every classical
solution to be an exterior region of a \rnads\ black hole with a
nondegenerate Killing horizon, with the spacelike  hypersurfaces
extending from the horizon bifurcation two-sphere to the asymptotically
\ads\ infinity. The constraints are simplified by a canonical
transformation, which generalizes that given by Kucha\v{r} in the
spherically symmetric vacuum Einstein theory, and the theory is reduced
to its true dynamical degrees of freedom. After quantization, the grand
partition function of a thermodynamical grand canonical ensemble is
obtained by analytically continuing the Lorentzian time evolution
operator to imaginary time and taking the trace. A~similar analysis
under slightly modified boundary conditions leads to the partition
function of a thermodynamical canonical ensemble. The thermodynamics in
each ensemble is analyzed, and the conditions that the (grand)
partition function be dominated by a classical Euclidean black hole
solution are found. When these conditions are satisfied, we recover in
particular the Bekenstein-Hawking entropy. The limit of a vanishing
cosmological constant is briefly discussed.
\end{abstract}

\narrowtext

\section{Introduction}
\label{sec:intro}

Hawking's celebrated result of black hole radiation \cite{hawkingCMP}
and related developments \cite{hh-vac,unruh1,israel1}
made it possible to consider thermodynamical equilibrium systems
involving black holes in the manner first anticipated by
Bekenstein \cite{bekenstein1,bekenstein2}. At a semiclassical,
``phenomenological," level, a black hole thermodynamical equilibrium
system can be introduced by simply immersing a radiating black hole in
a heat bath such that the outgoing Hawking radiation balances the
radiation that falls in from the
bath \cite{davies1,davies2,GP1,smolin1,gar-wald}.  At a deeper level,
one aspires to construct a full thermodynamical equilibrium ensemble by
starting from a quantum theory of gravity for black hole type
geometries \cite{GH1,hawkingCC,york1,BY-quasilocal,BYrev,cartei1}. For
reviews, see for example
Refs.\ \cite{davies2,BYrev,pagerev,wald-qft,carlip-rev}.

At the semiclassical level, the thermodynamical equilibrium
configurations involving black holes tend to be unstable against
thermal fluctuations \cite{davies1,davies2}.
The classic example is a Schwarzschild black hole in equilibrium with an
asymptotically flat heat bath, in the approximation where the
back-reaction of the radiation on the geometry is neglected: the heat
capacity in this instance is $-{(8\pi T^2)}^{-1}$, where $T$ is the
temperature measured at the infinity, and the fact that this heat
capacity is negative indicates thermodynamical instability. While such
instabilities are not unexpected in self-gravitating systems, they do
pose an obstacle to constructing thermodynamical equilibrium ensembles
from quantum gravity. This is because the existence of a thermodynamical
ensemble implies the positivity of certain response functions
associated with that ensemble \cite{reichl}. For example, in the
canonical ensemble the heat capacity is necessarily positive;
consequently, a canonical ensemble of the usual kind does not
appear to exist for Schwarzschild black holes in asymptotically flat
space \cite{sorkin-can}.

To construct a thermodynamical ensemble appropriate for black hole
geometries from a quantum theory of gravity, one thus needs to choose
the boundary conditions for the ensemble in a judicious manner,
motivated by the stability of the corresponding  semiclassical
equilibrium situations. One possibility is to replace an asymptotic
infinity by a finite ``box" at which the local temperature is then
fixed \cite{york1,BWY1,WYprl,whitingCQG,BBWY,%
B+,BMY,LW1,comer1,MW,LW2,lau1,BLPP}.
The possibility on which we shall concentrate in this paper is
to include a negative cosmological
constant \cite{pagerev,HPads,PagePhill,%
btz-cont,BrownCreMann,zaslavskii1}.

A~negative cosmological constant makes classical black hole solutions
asymptotically \ads. We shall consider spherically symmetric spacetimes,
and as the only matter field we include the spherically symmetric
Maxwell field. All the relevant classical solutions then belong to the
\rnads\ (RNAdS)
family \cite{hoffmann,carter-cmp,carter-les,exact-book}. The temperature
of the Hawking radiation is redshifted to zero at the asymptotically
\ads\ infinity, but from the rate at which the local Hawking temperature
approaches zero one can extract a ``renormalized" Hawking temperature,
and this renormalized Hawking temperature can then be taken as one
fixed quantity in the thermodynamical
ensembles \cite{pagerev,HPads,PagePhill,BrownCreMann}. We shall consider
both the canonical ensemble, in which the electric charge is fixed, and
the grand canonical ensemble, in which the electric potential
difference between the event horizon and the infinity is fixed.

To quantize the theory and to build the equilibrium ensembles, we shall
adapt the method introduced in Ref.\ \cite{LW2} in the context of
spherically symmetric vacuum geometries in the presence of a finite
boundary. We shall first set up a classical Lorentzian Hamiltonian
theory in which, on the classical solutions, the right end of the
spacelike hypersurfaces is at the asymptotically \ads\ infinity in an
exterior region of a black hole spacetime, and the left end of the
hypersurfaces is at the bifurcation two-sphere of a nondegenerate
Killing horizon. We then canonically quantize this theory, and obtain
the thermodynamical (grand) partition function by suitably continuing
the Schr\"odinger picture time evolution operator to imaginary time and
taking the trace. A~crucial input is how to handle the analytic
continuation at the bifurcation two-sphere.
As in Ref.\ \cite{LW2}, we shall see that a continuation motivated by
smoothness of Euclidean black hole geometries yields a (grand)
partition function that is in agreement with that obtained via path
integral methods.

To implement the method used in Ref.\ \cite{LW2}, one must be able to
canonically quantize the Lorentzian theory in some practical fashion.
In Ref.\ \cite{LW2} this was achieved by using canonical variables that
were first introduced by Kucha\v{r} under asymptotically flat,
Kruskal-like boundary conditions \cite{kuchar1}. In these variables the
constraints of the vacuum theory become exceedingly simple, and the
classical Hamiltonian theory can be explicitly reduced into an
unconstrained Hamiltonian theory with just one canonical pair of
degrees of freedom. We shall show that an analogous set of canonical
variables exists for our system, and the classical Hamiltonian theory
can again be explicitly reduced into an unconstrained Hamiltonian
theory. Under boundary conditions tailored to the grand
canonical ensemble, the reduced Lorentzian Hamiltonian theory has {\em
two\/} canonical pairs of degrees of freedom;\footnote{The conclusion
of two canonical pairs of degrees of freedom for the spherically
symmetric Einstein-Maxwell system with a cosmological constant was
previously reached, under a different set of boundary conditions, in
Ref.\ \cite{thiemann3}.}
under boundary conditions tailored to the canonical ensemble, the
reduced Lorentzian Hamiltonian theory has just one pair of canonical
degrees of freedom. Using these variables, it will be possible to
construct a quantum theory and a (grand) partition function in close
analogy with Ref.\ \cite{LW2}.

It will turn out that both the canonical ensemble and the grand
canonical ensemble for our system are well defined. In particular,
the appropriate thermodynamical response functions are positive.
We shall also be able to give the conditions under which the (grand)
partition function is dominated by a classical Euclidean solution. The
grand canonical ensemble exhibits a transition from a region where a
classical solution dominates to a region where no classical solution
dominates, in close analogy with what happens with the spherically
symmetric boxed vacuum canonical ensemble \cite{WYprl,whitingCQG}. As in
Refs.\ \cite{WYprl,whitingCQG}, one may see this as evidence for a phase
transition between a black hole sector and a topologically different
``hot \ads\ space" sector. In the canonical ensemble we find evidence
for this kind of a phase transition only in the special case when the
charge vanishes. When the charge is nonvanishing, there occurs a
different kind of phase transition in which the dominating contribution
to the partition function shifts from one classical solution to another
as the boundary data changes.

The rest of the paper is as follows. In Sec.\ \ref{sec:metric} we
set up a classical Hamiltonian theory under boundary conditions
tailored to the grand canonical thermodynamical ensemble, paying
special attention to the falloff conditions at the asymptotically \ads\
infinity \cite{Ash-ads,HennTeit-ads}. In particular, we choose to fix
the values of the electric potential at the infinity and at the
horizon, in a manner that will be made precise in terms of the
fall-off conditions. In Sec.\ \ref{sec:transformation} we perform the
canonical transformation, and in Sec.\ \ref{sec:reduction} the
constraints are eliminated and the theory is reduced to its true
dynamical degrees of freedom. In Sec.\ \ref{sec:quantum} we quantize
the theory and obtain the grand partition function of the
thermodynamical grand canonical ensemble. The thermodynamics in the
grand canonical ensemble is analyzed in
Sec.~\ref{sec:thermo-grand}\null. Sec.\
\ref{sec:can-ensemble} outlines the corresponding classical, quantum
mechanical, and thermodynamical analyses under boundary conditions that
fix the charge instead of the electric potential, and thus lead to the
thermodynamical canonical ensemble.

The results are summarized and discussed in
Sec.~\ref{sec:discussion}\null.  Some facts about the RNAdS solutions
are collected in Appendix~\ref{app:rnads}. Finally, Appendix
\ref{app:as-flat} outlines the classical Hamiltonian analysis and the
quantization of the reduced Hamiltonian theory in the case where the
cosmological constant vanishes and the asymptotically \ads\ falloff
conditions are replaced by asymptotically flat falloff conditions. With
asymptotic flatness, neither the partition function nor the grand
partition function turns out to be well-defined, and we recover neither
a canonical ensemble nor a grand canonical ensemble.

\section{Canonical formulation in the metric variables}
\label{sec:metric}

In this section we present a Hamiltonian formulation of spherically
symmetric electrovacuum spacetimes with a negative cosmological
constant, with boundary conditions appropriate for the exterior of a
RNAdS black hole with a nondegenerate horizon. Some relevant
properties of the RNAdS metric are reviewed in
Appendix~\ref{app:rnads}\null.

We consider the general spherically symmetric ADM metric
\begin{equation}
ds^2 = - N^2 dt^2 + \Lambda^2 {(dr + N^r dt)}^2 +R^2 d\Omega^2
\ \ ,
\label{4-metric}
\end{equation}
where $d\Omega^2$ is the metric on the unit two-sphere, and $N$,
$N^r$, $\Lambda$ and $R$ are functions of $t$ and $r$ only.
The electromagnetic potential is taken to be described by the
spherically symmetric one-form
\begin{equation}
A = \Gamma dr + \Phi dt
\ \ ,
\label{4-vectorpot}
\end{equation}
where $\Gamma$ and $\Phi$ are functions of $t$ and $r$ only.
The fact that this one-form is globally defined makes the
electromagnetic bundle trivial, and will preclude the black hole from
having a magnetic charge.
The coordinate $r$ takes the semi-infinite range $[0,\infty)$. Unless
otherwise stated, we shall assume both the spatial metric and the
spacetime metric to be nondegenerate. In particular, $\Lambda$, $R$,
and $N$ are taken to be positive. We shall work in natural units,
$\hbar = c = G = 1$.

The action of the Einstein-Maxwell theory with a negative
cosmological constant is
\begin{equation}
S = {1 \over 16\pi} \int d^4x \, \sqrt{-{}^{(4)}g} \,
\left(
{}^{(4)}R
+6 \ell^{-2}
- F^{\mu\nu} F_{\mu\nu}
\right)
\ \ + \ \ \hbox{(boundary terms)}
\ \ ,
\end{equation}
where ${}^{(4)}g$ is the determinant of the four-dimensional metric,
${}^{(4)}R$ is the four-dimensional Ricci scalar, and $F_{\mu\nu} =
\partial_\mu A_\nu - \partial_\nu A_\mu$ is the electromagnetic field
tensor. The cosmological constant has been written as $-3 \ell^{-2}$,
where $\ell>0$. Inserting the spherically symmetric fields
(\ref{4-metric}) and (\ref{4-vectorpot}) and integrating over the
two-sphere we obtain, up to boundary terms, the action
\begin{eqnarray}
S_\Sigma [\Lambda, R, \Gamma ; N, N^r, && \Phi]
\nonumber
\\
= \int dt \int_0^\infty dr \,
\bigg[
&&
-N^{-1}
\left(
R \bigl( - {\dot \Lambda}
+ (\Lambda N^r)' \bigr)
( - {\dot R} + R' N^r )
+ \case{1}{2} \Lambda
{( - {\dot R} + R' N^r )}^2
\right)
\nonumber
\\
&&
+ \casehalf N^{-1} \Lambda^{-1} R^2
{( {\dot \Gamma} - \Phi' )}^2
\nonumber
\\
&&
+ N
\left(
\Lambda^{-2} R R' \Lambda'
- \Lambda^{-1} R R''
- \case{1}{2} \Lambda^{-1} {R'}^{2}
+ \case{1}{2} \Lambda
+ \case{3}{2} \ell^{-2} \Lambda R^2
\right) \bigg]
\ \ .
\label{S-lag}
\end{eqnarray}
The equations of motion derived from local variations of
(\ref{S-lag}) are the full Einstein-Maxwell equations for the
spherically symmetric fields (\ref{4-metric}) and~(\ref{4-vectorpot}).
A generalized Birkhoff's theorem can be proven using the same
techniques as in the case of a vanishing cosmological
constant \cite{MTW-birk}: every classical solution is locally either a
member of the extended RNAdS family (see Appendix~\ref{app:rnads}), or
a spacetime that generalizes the Bertotti-Robinson
solution to accommodate a negative cosmological
constant \cite{carter-cmp,carter-les,exact-book,MTW-birk}. We shall
address the boundary conditions and boundary terms that are needed to
make the variational principle globally well-defined after passing to
the Hamiltonian formulation.

The momenta conjugate to the configuration variables $\Lambda$, $R$,
and $\Gamma$ are
\begin{mathletters}
\begin{eqnarray}
P_{\Lambda} &=& - N^{-1} R ( {\dot R} - R' N^r )
\ \ ,
\label{PLambda}
\\
P_R &=& -N^{-1}
\left( \Lambda
( {\dot R} - R' N^r )
+ R \bigl(
{\dot \Lambda} - (\Lambda N^r)'
\bigr)
\right)
\ \ ,
\label{PR}
\\
P_\Gamma &=& N^{-1} \Lambda^{-1} R^2
( {\dot \Gamma} - \Phi' )
\ \ .
\label{PB}
\end{eqnarray}
\end{mathletters}%
A~Legendre transformation gives the Hamiltonian action
\begin{eqnarray}
S_\Sigma && [\Lambda, R, \Gamma, P_\Lambda, P_R, P_\Gamma ;
N, N^r, {\tilde\Phi}]
\nonumber
\\
&&= \int dt
\int_0^\infty dr \left( P_\Lambda {\dot \Lambda} +
P_R {\dot R} + P_\Gamma {\dot \Gamma}
- NH - N^r H_r - {\tilde \Phi} G
\right)
\ \ ,
\label{S-ham}
\end{eqnarray}
where the super-Hamiltonian constraint~$H$, the radial supermomentum
constraint~$H_r$, and the Gauss law constraint $G$ are given by
\begin{mathletters}
\label{constraints}
\begin{eqnarray}
H &=& - R^{-1} P_R P_\Lambda
+ \case{1}{2} R^{-2} \Lambda
( P_\Lambda^2 + P_\Gamma^2)
\nonumber
\\
&&
+ \Lambda^{-1} R R'' - \Lambda^{-2} R R' \Lambda'
+ \case{1}{2} \Lambda ^{-1} {R'}^2
- \case{1}{2} \Lambda
- \case{3}{2} \ell^{-2} \Lambda R^2
\ \ ,
\label{superham}
\\
H_r &=& P_R R' - \Lambda P_\Lambda' - \Gamma P_\Gamma'
\ \ ,
\label{supermom}
\\
G &=& - P_\Gamma'
\ \ .
\label{gauss}
\end{eqnarray}
\end{mathletters}%
We have written the electric potential $\Phi$ in terms of the quantity
\begin{equation}
{\tilde \Phi} := \Phi - N^r \Gamma
\ \ ,
\label{Phi-redef}
\end{equation}
which now acts as the Lagrange multiplier associated with the Gauss
constraint in~(\ref{S-ham}). It would be possible to proceed
retaining $\Phi$ as the Lagrange multiplier, and the
supermomentum constraint would then be the same as without the
electromagnetic field
(see, for example, Ref.\ \cite{kraus-wilczek2}).
However, using ${\tilde \Phi}$ has the technical advantage that the
supermomentum constraint (\ref{supermom}) generates spatial
diffeomorphisms in both the gravitational and electromagnetic
variables. This fact will prove useful in
Sec.~\ref{sec:transformation}\null.

The Hamiltonian equations of motion are obtained from local
variations of~(\ref{S-ham}). The constraint equations are
\begin{mathletters}
\begin{eqnarray}
H &=& 0
\ \ ,
\\
H_r &=& 0
\ \ ,
\\
G &=& 0
\ \ ,
\label{constr-eqs-G}
\end{eqnarray}
\end{mathletters}%
and the dynamical equations of motion read
\begin{mathletters}
\begin{eqnarray}
{\dot \Lambda}
&=&
N ( R^{-2} \Lambda P_\Lambda - R^{-1} P_R )
+ {\left( N^r \Lambda \right)}'
\ \ ,
\\
{\dot R}
&=&
- N R^{-1} P_\Lambda
+ N^r R'
\ \ ,
\\
{\dot \Gamma}
&=&
N \Lambda R^{-2} P_\Gamma
+ {\left( N^r \Gamma \right)}'
+ {\tilde \Phi}'
\ \ ,
\\
{\dot P}_\Lambda
&=&
\casehalf N
\left[
- R^{-2} ( P_\Lambda^2 + P_\Gamma^2 )
- {(\Lambda^{-1} R')}^2 + 1 + 3
\ell^{-2} R^2 \right]
- \Lambda^{-2} N' R R' + N^r P_\Lambda'
\ \ ,
\\
{\dot P}_R
&=&
N
\left[
\Lambda R^{-3} ( P_\Lambda^2 + P_\Gamma^2 ) - R^{-2} P_\Lambda P_R -
{(\Lambda^{-1} R')}' + 3 \ell^{-2} \Lambda R \right]
\nonumber
\\
&&
\
- {( \Lambda^{-1} N' R )}'
 + {\left( N^r P_R \right)}'
\ \ ,
\\
{\dot P}_\Gamma
&=&
N^r P_\Gamma'
\ \ .
\label{dyn-eqs-PB}
\end{eqnarray}
\end{mathletters}%
It is easy to verify that the Poisson bracket algebra of the constraints
closes, and we thus have a first class constrained
system \cite{henn-teit-book}.

We now wish to adopt boundary conditions that enforce every classical
solution to be an exterior region of a RNAdS spacetime with
a nondegenerate horizon (see Appendix~\ref{app:rnads}),
such that the constant $t$ hypersurfaces begin at the horizon
bifurcation two-sphere at $r=0$ and reach the asymptotically \ads\
infinity as $r\to\infty$.

Consider first the left end of the hypersurfaces. At $r\to0$, we
adopt the conditions
\begin{mathletters}
\label{s-r}
\begin{eqnarray}
\Lambda (t,r) &=& \Lambda_0(t) + O(r^2)
\ \ ,
\label{s-r-Lambda}
\\
R(t,r) &=& R_0(t) + R_2(t) r^2 + O(r^4)
\ \ ,
\label{s-r-R}
\\
P_{\Lambda}(t,r) &=& O(r^3)
\ \ ,
\label{s-r-PLambda}
\\
P_{R}(t,r) &=& O(r)
\ \ ,
\label{s-r-PR}
\\
N(t,r) &=& N_1(t)r + O(r^3)
\ \ ,
\label{s-r-N}
\\
N^r(t,r) &=& N^r_1(t)r + O(r^3)
\ \ ,
\label{s-r-Nr}
\\
\Gamma(t,r) &=& O(r)
\ \ ,
\label{s-r-B}
\\
P_\Gamma(t,r) &=& Q_0(t) + Q_2(t) r^2 + O(r^4)
\ \ ,
\label{s-r-PB}
\\
{\tilde \Phi}(t,r) &=& {\tilde \Phi}_0(t) + O(r^2)
\ \ ,
\label{s-r-tPhi}
\end{eqnarray}
\end{mathletters}%
where $\Lambda_0$ and $R_0$ are positive, and $N_1\ge0$. Here
$O(r^n)$ stands for a term whose magnitude at $r\to0$ is bounded by
$r^n$ times a constant, and whose $k$'th derivative at $r\to0$ is
similarly bounded by $r^{n-k}$ times a constant for $1\le k\le n$.
It is straightforward to verify that these falloff conditions are
consistent with the constraints $H=H_r=G=0$, and that they are
preserved by the time evolution equations. The metric falloff
conditions (\ref{s-r-Lambda})--(\ref{s-r-Nr}), which are identical to
those introduced in Ref.\ \cite{LW2} in the context of the
Schwarzschild black hole, guarantee that the classical solutions have
a nondegenerate horizon, and that the constant $t$ hypersurfaces
begin at $r=0$ at a horizon bifurcation two-sphere in a manner
asymptotic to hypersurfaces of constant Killing time.\footnote{The
text in Ref.\ \cite{LW2} contains at this point a minor inaccuracy.
Equations (2.6a) and (2.6b) of Ref.\ \cite{LW2} [our (\ref{s-r-Lambda})
and (\ref{s-r-R})] are not sufficient to ensure that the
hypersurfaces end at the horizon bifurcation two-sphere, but for example
the set (2.6a)--(2.6c) of Ref.\ \cite{LW2} [our
(\ref{s-r-Lambda})--(\ref{s-r-PLambda})] is.}
The coordinates become thus singular at $r\to0$, but this singularity is
quite precisely controlled. In particular, on a classical solution the
future unit normal to a constant $t$ hypersurface defines at $r\to0$ a
future timelike unit vector $n^a(t)$ at the bifurcation two-sphere, and
the evolution of the constant $t$ hypersurfaces boosts this vector
according to
\begin{equation}
n^a(t_1) n_a(t_2) =
-\cosh\left(\int_{t_1}^{t_2}
\Lambda_0^{-1}(t) N_1(t) \, dt \right)
\ \ .
\label{n-boost}
\end{equation}
The falloff conditions (\ref{s-r-B})--(\ref{s-r-tPhi}) for the
electromagnetic field variables are motivated by our thermodynamical
goal, and they will be discussed further in
Sec.~\ref{sec:quantum}\null.

Consider then the right end of the hypersurfaces. At $r\to\infty$, we
assume that the variables have asymptotic expansions in integer
powers of $(1/r)$, with the leading order behavior
\begin{mathletters}
\label{l-r}
\begin{eqnarray}
\Lambda(t,r)
&=&
\ell r^{-1} - \casehalf \ell^3 r^{-3} + \lambda(t) \ell^3 r^{-4}
+ O^\infty (r^{-5})
\ \ ,
\label{l-r-Lambda}
\\
R(t,r)
&=&
r + \ell^2 \rho(t) r^{-2} + O^\infty (r^{-3})
\ \ ,
\label{l-r-R}
\\
P_{\Lambda}(t,r) &=& O^\infty (r^{-2})
\ \ ,
\label{l-r-PLambda}
\\
P_{R}(t,r) &=& O^\infty (r^{-4})
\ \ ,
\label{l-r-PR}
\\
N(t,r) &=& \Lambda^{-1} R' \left( {\tilde N}_+(t) + O^\infty(r^{-5})
\right)
\ \ ,
\label{l-r-N}
\\
N^r(t,r) &=& O^\infty (r^{-2})
\ \ ,
\label{l-r-Nr}
\\
\Gamma(t,r) &=& O^\infty (r^{-2})
\ \ ,
\label{l-r-B}
\\
P_\Gamma(t,r) &=& Q_+(t) + O^\infty (r^{-1})
\ \ ,
\label{l-r-PB}
\\
{\tilde \Phi}(t,r) &=& {\tilde \Phi}_+(t) + O^\infty (r^{-1})
\ \ ,
\label{l-r-tPhi}
\end{eqnarray}
\end{mathletters}%
where ${\tilde N}_+(t)>0$. $O^\infty(r^{-n})$ denotes a term that
falls off at infinity as~$r^{-n}$, and whose derivatives with respect
to $r$ fall off accordingly as~$r^{-n-k}$, $k=1,2,\ldots$. It is
again straightforward to verify that these falloff conditions are
consistent with the constraints and that they are preserved by the
time evolution equations. Comparison with Ref.\ \cite{HennTeit-ads}
shows that the metric is asymptotically \ads, with the constant $t$
hypersurfaces being asymptotic to hypersurfaces of constant Killing
time, and ${\tilde N}_+(t)$ gives the rate at which the Killing time
evolves with respect to $t$ at the infinity. Note that the
lapse-function $N$ diverges at the infinity for any nonzero value
of~${\tilde N}(t)$. For future use, we define the quantity
\begin{equation}
M_+(t) := \lambda(t) + 3\rho(t)
\ \ .
\label{M+def}
\end{equation}
When the equations of motion hold, $M_+(t)$ is independent of~$t$, and
it is equal to the mass parameter of the RNAdS
metric~(\ref{rnads-metric}).

Taken together, the falloff conditions (\ref{s-r}) and (\ref{l-r})
achieve our aim.  Every classical solution is an exterior region of a
RNAdS spacetime with a nondegenerate event horizon, such that the
constant $t$ hypersurfaces begin at the horizon bifurcation two-sphere
and reach the asymptotically \ads\ infinity. In particular, the
classical solutions satisfy $R_2>0$.\footnote{The falloff conditions
(\ref{s-r}) are compatible with either sign of~$R_2$. The case $R_2<0$
would correspond to the bifurcation two-sphere of an inner horizon,
which is excluded from the classical solutions only after the
asymptotically \ads\ falloff has been invoked at $r\to\infty$. If
desired, the requirement $R_2>0$ could of course be already added to
the conditions~(\ref{s-r}).}

It would be possible to replace (\ref{l-r-Lambda}) and (\ref{l-r-R})
by
\begin{mathletters}
\label{l-r-prime}
\begin{eqnarray}
\Lambda(t,r)
&=&
\ell r^{-1} - \casehalf \ell^3 r^{-3} + O^\infty (r^{-4})
\ \ ,
\label{l-r-Lambda-prime}
\\
R(t,r)
&=&
r + O^\infty (r^{-2})
\ \ ,
\label{l-r-R-prime}
\end{eqnarray}
\end{mathletters}%
and then drop the assumption that the expansion proceed in integer
powers of $(1/r)$ beyond the order shown, provided one makes more
precise assumptions about what is meant by the symbol $O^\infty$.
Alternatively, it would be possible to strengthen the falloff
conditions to read
\begin{mathletters}
\label{l-r-pprime}
\begin{eqnarray}
\Lambda(t,r)
&=&
\ell r^{-1} - \casehalf \ell^3 r^{-3} + \lambda(t) \ell^3 r^{-4}
+ O^\infty (r^{-4-\epsilon})
\ \ ,
\label{l-r-Lambda-pprime}
\\
R(t,r)
&=&
r + O^\infty (r^{-2-\epsilon})
\ \ ,
\label{l-r-R-pprime}
\end{eqnarray}
\end{mathletters}%
where $0<\epsilon\le1$, with similar changes in the rest
of~(\ref{l-r}). This would be analogous to the falloff conditions
adopted for the asymptotically flat Schwarzschild case in
Ref.\ \cite{kuchar1},
in that the value of the mass (\ref{M+def}) could
then be read solely from the expansion (\ref{l-r-Lambda-pprime})
of~$\Lambda$. One might also consider writing the theory in terms of a
lapse-function that has been rescaled by the factor $\Lambda^{-1}R'$:
by~(\ref{l-r-N}), the falloff of the new lapse at $r\to\infty$ would
then be independent of the canonical variables. For concreteness, we
shall adhere to the theory as written above.

We can now write an action principle compatible with our
falloff conditions. Consider the total action
\begin{eqnarray}
S [\Lambda, R, \Gamma, P_\Lambda, P_R, P_\Gamma ;
N, N^r, {\tilde \Phi}]
&=&
S_\Sigma [\Lambda, R, \Gamma, P_\Lambda, P_R, P_\Gamma ;
N, N^r, {\tilde\Phi}]
\nonumber
\\
&&+ S_{\partial\Sigma} [\Lambda, R, Q_0, Q_+ ; N, {\tilde \Phi}_0,
{\tilde \Phi}_+]
\ \ ,
\label{S-total}
\end{eqnarray}
where the boundary action is
\begin{eqnarray}
S_{\partial\Sigma} && [\Lambda, R, Q_0, Q_+ ; N, {\tilde \Phi}_0,
{\tilde \Phi}_+]
\nonumber
\\
&&=
\int dt \left(
\casehalf R_0^2 N_1 \Lambda_0^{-1}
- {\tilde N}_+ M_+
+ {\tilde \Phi}_0 Q_0
- {\tilde \Phi}_+ Q_+ \right)
\ \ .
\label{S-boundary}
\end{eqnarray}
The total action (\ref{S-total}) is clearly well-defined under our
boundary conditions. Its variation contains a volume term
proportional to the equations of motion, boundary terms from the
initial and final hypersurfaces proportional to $\delta \Lambda$,
$\delta R$, and~$\delta \Gamma$, and boundary terms from $r=0$ and
$r=\infty$ given by
\begin{equation}
\int dt \, \left(
\casehalf R_0^2 \,
\delta \! \left( N_1
\Lambda_0^{-1} \right)
- M_+ \, \delta {\tilde N_+}
+ Q_0 \, \delta {\tilde \Phi}_0
- Q_+ \, \delta {\tilde \Phi}_+
\right)
\ \ .
\label{bt-0-infty}
\end{equation}
The variation thus gives the desired classical equations of motion
provided we fix, in addition to the initial and final values of
$\Lambda$, $R$, and~$\Gamma$, also the quantities $N_1\Lambda_0^{-1}$,
${\tilde N_+}$, ${\tilde \Phi}_0$, and~${\tilde \Phi}_+$. On a
classical solution all these quantities have a clear geometrical
interpretation. $N_1\Lambda_0^{-1}$ gives via (\ref{n-boost}) the
evolution of the unit normal to the constant $t$  hypersurface at the
bifurcation two-sphere, and ${\tilde N_+}$ gives the evolution of the
Killing time at the infinity. ${\tilde \Phi}_0$~and
${\tilde\Phi}_+$ describe the electromagnetic gauge in a way that will
become more transparent in Sec.~\ref{sec:reduction}\null. Note from
(\ref{s-r-tPhi}) that when a classical solution is written in
coordinates that are regular at the bifurcation two-sphere, the
electromagnetic potential will be regular at the bifurcation two-sphere
only if ${\tilde \Phi}_0=0$.

\section{Canonical transformation}
\label{sec:transformation}

In this section we perform a canonical transformation, which
generalizes that given in Ref.\ \cite{kuchar1} for the spherically
symmetric vacuum Einstein theory. Following Ref.\ \cite{kuchar1}, we
first examine how the variables appearing in the action (\ref{S-total})
carry the information about the geometry of the classical
solution~(\ref{rnads-metric}). We then use this information as a guide
for finding the canonical transformation.

\subsection{Reconstruction}
\label{subsec:reconstruction}

Under our boundary conditions, every classical solution is an
exterior region of a RNAdS spacetime with a nondegenerate Killing
horizon (see Appendix~\ref{app:rnads}). We now assume that we are
given the canonical data $( \Lambda, R, \Gamma, P_\Lambda, P_R,
P_\Gamma)$ on a spacelike hypersurface embedded in such a RNAdS
spacetime. We wish to recover from the canonical data the mass and
charge parameters of the spacetime, the information about the embedding
of the  hypersurface in the spacetime, and the information about the
electromagnetic gauge.

Consider first the charge. The equations of motion imply that
$P_\Gamma$ is independent of both $t$ and~$r$. It is easily seen that
in the curvature coordinates~(\ref{rnads-metric}), the value of
$P_\Gamma$ is just the charge~$Q$. As $P_\Gamma$ is unchanged by the
gauge transformations generated by the constraints, it follows that in
any gauge
\begin{equation}
Q = P_\Gamma
\ \ .
\label{Q=PB}
\end{equation}

Consider then the mass. The reconstruction of the function $F$
appearing in the metric (\ref{curv-metric}) proceeds exactly as in
Ref.\ \cite{kuchar1}, with the result
\begin{equation}
F = {\left( {R' \over \Lambda} \right)}^2
- {\left( { P_\Lambda \over R} \right)}^2
\ \ .
\label{F-def}
\end{equation}
{}From (\ref{rnads-F}) and~(\ref{Q=PB}), we find for the mass
the expression
\begin{equation}
M = {R \over 2} \left( {R^2 \over \ell^2} + 1 +
{ P_\Gamma^2 \over R^2}
- F \right)
\ \ ,
\label{M-class}
\end{equation}
where $F$ is understood to be given by~(\ref{F-def}).

Consider then the embedding. By repeating the steps in
Ref.\ \cite{kuchar1}, we obtain
\begin{equation}
-T' = R^{-1} F^{-1} \Lambda P_\Lambda
\ \ ,
\label{Tprime}
\end{equation}
which determines the embedding up to an overall additive constant
in~$T$. To determine the value of the additive constant, one needs to
know the value of $T$ at one point on the hypersurface.

Consider finally the electromagnetic gauge. By~(\ref{rnads-A}), there
exists a function $\xi(t,r)$ such that
\begin{eqnarray}
A &=& R^{-1} Q dT + d\xi
\nonumber
\\
&=&
\left( R^{-1} Q T' + \xi' \right) dr
+
\left( R^{-1} Q {\dot T} + {\dot \xi} \right) dt
\ \ .
\label{Xi-def}
\end{eqnarray}
{}From~(\ref{4-vectorpot}), (\ref{Q=PB}), and (\ref{Tprime}) we then
obtain
\begin{equation}
\xi' = \Gamma + R^{-2} F^{-1} \Lambda P_\Lambda P_\Gamma
\ \ ,
\label{Xiprime}
\end{equation}
which determines the value of $\xi$ on the hypersurface up to an
additive constant.

\subsection{Transformation}
\label{subsec:transformation}

We have seen that when the equations of motion hold, the quantities
defined by Eqs.\ (\ref{Q=PB})--(\ref{Tprime}) have a transparent
geometrical meaning. We now promote these equations into definitions of
functions on the phase space, valid even when the equations of motion
do not hold. Our aim is to complete the set of functions into a set that
constitutes a canonical chart.

We shall from now on assume that the quantity $R_2$ in
Eq.~(\ref{s-r-R}) is positive. As noted in Sec.~\ref{sec:metric},
this is always the case for our classical solutions.

The functions $M$ (\ref{M-class}) and $Q$ (\ref{Q=PB}) Poisson
commute with each other. The function $-T'$ (\ref{Tprime}) Poisson
commutes with $Q$ and is canonically conjugate to~$M$. This suggests
looking for a canonical transformation such that $M$ and $Q$ become
two new coordinates, and $-T'$ becomes the momentum conjugate to~$M$.
As in the Schwarzschild case \cite{kuchar1}, the function ${\sf R}:=R$
Poisson commutes with $M$, $Q$, and~$-T'$, and provides therefore a
candidate for a new canonical coordinate. The crucial issue then is
whether one can find momenta conjugate to $Q$ and ${\sf R}$ such that
the transformation is canonical.

A~necessary condition for the prospective new momenta arises from the
observation that the supermomentum constraint (\ref{supermom})
generates spatial diffeomorphisms in all the variables. Since $M$,
$Q$, and ${\sf R}$ are spatial scalars, the expression for $H_r$ in the
new variables must be $P_M M' + P_Q Q' + P_{\sf R} {\sf R}'$. Equating
this with (\ref{supermom}) and substituting for $M$ and $P_M=-T'$
their expressions from (\ref{M-class}) and (\ref{Tprime}) gives only
one equation for the two unknowns $P_Q$ and~$P_{\sf R}$, but the
structure of the equation as a linear combination of $R'$ and
$P_\Gamma'$ suggests setting the coefficients of $R'$ and $P_\Gamma'$
individually to zero. These considerations suggest the transformation
\begin{mathletters}
\label{trans}
\begin{eqnarray}
M &:=& \casehalf R \left( R^2 \ell^{-2} + 1 + P_\Gamma^2 R^{-2} - F
\right)
\ \ ,
\label{trans-M}
\\
P_M &:=& R^{-1} F^{-1} \Lambda P_\Lambda
\ \ ,
\label{trans-PM}
\\
{\sf R} &:=& R
\ \ ,
\label{trans-sfR}
\\
P_{\sf R} &:=&
P_R
- \case{1}{2} R^{-1} \Lambda P_\Lambda
- \case{1}{2} R^{-1} F^{-1} \Lambda P_\Lambda
\nonumber
\\
&& \; - R^{-1} \Lambda^{-2} F^{-1}
\left(
{(\Lambda P_\Lambda)}' (RR')
- (\Lambda P_\Lambda) {(RR')}'
\right)
\nonumber
\\
&& \; + \casehalf R^{-1} F^{-1} \Lambda P_\Lambda
\left( P_\Gamma^2 R^{-2} - 3 R^3 \ell^{-2} \right)
\ \ ,
\label{trans-PsfR}
\\
Q &:=& P_\Gamma
\ \ ,
\label{trans-Q}
\\
P_Q &:=& -\Gamma - R^{-2} F^{-1} \Lambda P_\Lambda P_\Gamma
\ \ ,
\label{trans-PQ}
\end{eqnarray}
\end{mathletters}%
where $F$ is given by~(\ref{F-def}). The analogy between the pairs
$(M, P_M)$ and $(Q, P_Q)$ becomes manifest by observing from
(\ref{Xiprime}) that on a classical solution, $P_Q$~carries the
information about the electromagnetic gauge via $P_Q=-\xi'$.

It is now straightforward to demonstrate that the transformation
(\ref{trans}) is indeed canonical. We begin with the identity
\begin{eqnarray}
P_\Lambda \delta \Lambda
&&
+ P_R \delta R
+ P_\Gamma \delta \Gamma
- P_M \delta M
- P_{\sf R} \delta {\sf R}
- P_Q \delta Q
\nonumber
\\
&&
=
{\left( \case{1}{2} R \delta R
\ln \left|
{RR'+ \Lambda P_\Lambda \over RR'
- \Lambda P_\Lambda} \right|
\vphantom{
{\left|
{RR'+ \Lambda P_\Lambda \over RR'
- \Lambda P_\Lambda} \right|}^A_A
}
\right)}'
\; + \;
\delta
\left( \Gamma P_\Gamma + \Lambda P_\Lambda
+ \case{1}{2} RR' \ln \left|
{RR' - \Lambda P_\Lambda \over RR'
+ \Lambda P_\Lambda}
\right|
\vphantom{
{\left|
{RR'+ \Lambda P_\Lambda \over RR'
- \Lambda P_\Lambda} \right|}^A_A
}
\right)
\ \ ,
\label{diff-PdQ}
\end{eqnarray}
and integrate both sides with respect to $r$ from $r=0$ to
$r=\infty$. The first term on the right hand side gives substitution
terms from $r=0$ to $r=\infty$ that vanish by virtue of our falloff
conditions, and we obtain
\begin{eqnarray}
\int_0^\infty && dr
\left( P_\Lambda \delta \Lambda + P_R \delta R
+ P_\Gamma \delta \Gamma
\right)
\nonumber
\\
&&-\int_0^\infty dr \left( P_M \delta M + P_{\sf R} \delta {\sf R} +
P_Q \delta Q
\right)
= \delta \omega
\left[
\Lambda, R, \Gamma, P_\Lambda, P_\Gamma
\right]
\ \ ,
\label{diff-liouville}
\end{eqnarray}
where
\begin{equation}
\omega
\left[
\Lambda, R, \Gamma, P_\Lambda, P_\Gamma
\right]
=
\int_0^\infty dr
\left( \Gamma P_\Gamma + \Lambda P_\Lambda
+ \case{1}{2} RR' \ln \left|
{RR' - \Lambda P_\Lambda \over RR'
+ \Lambda P_\Lambda}
\right|
\vphantom{
{\left|
{RR'+ \Lambda P_\Lambda \over RR'
- \Lambda P_\Lambda} \right|}^A_A
}
\right)
\ \ .
\label{omega}
\end{equation}
The functional $\omega \left[ \Lambda, R, \Gamma, P_\Lambda, P_\Gamma
\right]$ is well-defined by virtue of the falloff conditions.
Eqs.~(\ref{diff-liouville}) and (\ref{omega}) show that the Liouville
forms of the old and new variables differ only by an exact form, and
the transformation is thus canonical.

The new variables have well-defined falloff properties at $r=0$ and
$r\to\infty$. At $r=0$, Eqs.~(\ref{s-r}) imply
\begin{equation}
F(t,r) = 4 R_2^2 \Lambda_0^{-2} r^2 + O(r^4)
\label{sF2-r}
\end{equation}
and
\begin{mathletters}
\label{s2-r}
\begin{eqnarray}
M (t,r) &=& M_0(t) + M_2(t) r^2 + O(r^4)
\ \ ,
\label{s2-r-M}
\\
{\sf R}(t,r) &=& R_0(t) + R_2(t) r^2 + O(r^4)
\ \ ,
\\
Q (t,r) &=& Q_0(t) + Q_2(t) r^2 + O(r^4)
\ \ ,
\label{s2-r-Q}
\\
P_M(t,r) &=& O(r)
\ \ ,
\\
P_{\sf R} (t,r) &=& O(r)
\ \ ,
\\
P_Q (t,r) &=& O(r)
\ \ ,
\end{eqnarray}
\end{mathletters}
where
\begin{mathletters}
\begin{eqnarray}
M_0 &=&
\casehalf R_0 \left( R_0^2 \ell^{-2} + 1 + Q_0^2 R_0^{-2} \right)
\ \ ,
\label{M0}
\\
M_2 &=&
\casehalf R_2
\left( 3 R_0^2 \ell^{-2} + 1 - Q_0^2 R_0^{-2} - 4 R_0 R_2
\Lambda_0^{-2} \right)
+ Q_0 Q_2 R_0^{-1}
\ \ .
\label{M2}
\end{eqnarray}
\end{mathletters}
At $r\to\infty$, Eqs.~(\ref{l-r}) imply
\begin{mathletters}
\label{l2-r}
\begin{eqnarray}
M (t,r) &=& M_+(t) + O^\infty(r^{-1})
\ \ ,
\\
{\sf R}(t,r) &=& r + \ell^2 \rho(t) r^{-2} + O^\infty (r^{-3})
\ \ ,
\\
Q (t,r) &=& Q_+(t) + O(r^{-1})
\ \ ,
\\
P_M(t,r) &=& O^\infty (r^{-6})
\ \ ,
\\
P_{\sf R} (t,r) &=& O^\infty (r^{-4})
\ \ ,
\\
P_Q (t,r) &=& O^\infty (r^{-2})
\ \ ,
\end{eqnarray}
\end{mathletters}%
where $M_+(t)$ is given by~(\ref{M+def}).

The canonical transformation (\ref{trans}) becomes singular
when $F=0$. Under our boundary conditions the classical solutions
have $F>0$ for $r>0$.
At the limit $r\to0$ $F$ approaches zero according
to~(\ref{sF2-r}), but (\ref{trans}) has still a
well-defined limit obeying~(\ref{s2-r}).
Our canonical transformation is therefore well-defined
and differentiable near the classical solutions,
and similarly the inverse transformation is
well-defined and differentiable near the classical
solutions. {}From now on we shall assume that we are
always in a neighborhood of the classical
solutions such that $F>0$ holds for $r>0$.

\subsection{Action}
\label{subsec:action}

It is possible to write an action in the new variables by simply
re-expressing the constraints (\ref{constraints}) in terms of the new
coordinates and momenta. A~more transparent action can be
found if we exercise the freedom to redefine the Lagrange
multipliers.

The constraint terms in the bulk action (\ref{S-ham}) take the form
\begin{equation}
NH + N^r H_r + {\tilde \Phi} G
=
N^M M' + N^{\sf R} P_{\sf R} + N^Q Q'
\ \ ,
\label{constr-trans1}
\end{equation}
where
\begin{mathletters}
\label{N-def1}
\begin{eqnarray}
N^M &=&
- N F^{-1} \Lambda^{-1} R'
+ N^r R^{-1} F^{-1} \Lambda P_\Lambda
\ \ ,
\\
N^{\sf R} &=&
- N R^{-1} P_\Lambda
+ N^r R'
\ \ ,
\\
N^Q &=&
N R^{-1} F^{-1} \Lambda^{-1} R' P_\Gamma
- N^r \left( \Gamma + R^{-2} F^{-1} \Lambda P_\Lambda P_\Gamma \right)
- {\tilde \Phi}
\ \ .
\end{eqnarray}
\end{mathletters}%
When viewed as a linear transformation from $(N, N^r, {\tilde \Phi})$
to $(N^M, N^{\sf R}, N^Q)$, Eqs.\ (\ref{N-def1}) are nonsingular
for $r>0$. This suggests that we could take the constraint terms in
the new bulk action to be those on the right hand side
of~(\ref{constr-trans1}), with $N^M$, $N^{\sf R}$, and $N^Q$ as
independent Lagrange multipliers. At $r\to\infty$ this would be
satisfactory: (\ref{N-def1})~implies the asymptotic behavior
\begin{mathletters}
\label{N-l-1}
\begin{eqnarray}
N^M (t,r)
&=&
- {\tilde N}_+(t) + O^\infty(r^{-5})
\ \ ,
\\
N^{\sf R} (t,r)
&=&
O^\infty(r^{-2})
\ \ ,
\label{N-l-1-R}
\\
N^Q (t,r)
&=&
- {\tilde \Phi}_+(t)
+ O^\infty(r^{-1})
\ \ ,
\end{eqnarray}
\end{mathletters}%
and one could then fix ${\tilde N}_+(t)$ and ${\tilde \Phi}_+(t)$ as
in Sec.\ \ref{sec:metric} after adding the boundary action
\begin{equation}
- \int dt \left(
{\tilde N}_+ M_+
+ {\tilde \Phi}_+ Q_+ \right)
\ \ .
\end{equation}
However, at $r=0$ we have
\begin{mathletters}
\label{N-s-1}
\begin{eqnarray}
N^M (t,r)
&=&
- \casehalf N_1 \Lambda_0 R_2^{-1}
+ O(r^2)
\ \ ,
\\
N^{\sf R} (t,r)
&=&
O(r^2)
\ \ ,
\label{N-s-1-R}
\\
N^Q (t,r)
&=&
- {\tilde \Phi}_0 (t)
+ \casehalf N_1 \Lambda_0 Q_0 R_2^{-1} R_0^{-1}
+ O(r^2)
\ \ ,
\end{eqnarray}
\end{mathletters}%
which says that fixing $N^M$ and $N^Q$ at $r=0$ to values that are
independent of the canonical variables is not equivalent to fixing
$N_1\Lambda_0^{-1}$ and ${\tilde \Phi}_0$ to values that are
independent of the canonical variables. We therefore need to redefine
$N^M$ and $N^Q$ near $r=0$, without affecting their behavior at
$r\to\infty$.

To proceed, we make two assumptions. First, we assume $M_0>M_{\rm
crit}(Q_0)$, where  the function $M_{\rm crit}$ is defined in
Appendix~\ref{app:rnads}\null. Second, we regard Eq.\ (\ref{M0})
as defining $R_0$ in terms of $M_0$ and $Q_0$ as $R_0 = R_{\rm
hor}(M_0,Q_0)$, where the function $R_{\rm hor}$ is defined in
Appendix~\ref{app:rnads}\null. As discussed in Appendix~\ref{app:rnads},
these assumptions are always true for our classical solutions, and
they therefore merely tighten the neighborhood of the classical
solutions in which the field variables may take values. For future use,
we note that these assumptions imply $3 R_0^2 \ell^{-2} + 1 - Q_0^2
R_0^{-2}>0$, and the variation of $R_0$ takes the form
\begin{equation}
\delta R_0 =
2 {\left(3 R_0^2 \ell^{-2} + 1 - Q_0^2 R_0^{-2}\right)}^{-1}
\left(
\delta M_0 - R_0^{-1} Q_0 \delta Q_0
\right)
\ \ .
\label{deltaR0}
\end{equation}

Define now the quantities ${\tilde N}^M$ and ${\tilde N}^Q$ by
\begin{mathletters}
\label{N-def2}
\begin{eqnarray}
N^M
&=&
- {\tilde N}^M
\left[
(1-g)
+
2 g R_0
{\left(3 R_0^2 \ell^{-2} + 1 - Q_0^2 R_0^{-2}\right)}^{-1}
\right]
\ \ ,
\\
N^Q
&=&
2 {\tilde N}^M
g
Q_0
{\left(3 R_0^2 \ell^{-2} + 1 - Q_0^2 R_0^{-2}\right)}^{-1}
-
{\tilde N}^Q
\ \ ,
\end{eqnarray}
\end{mathletters}%
where $g(r)$ is a smooth decreasing function that vanishes at
$r\to\infty$ as $O^\infty(r^{-5})$, and approaches the value $1$ at
$r\to0$ as $g(r) = 1 + O(r^2)$. Eqs.\ (\ref{N-def2}) then define
a nonsingular linear transformation from $(N^M, N^{\sf R})$ to
$({\tilde N}^M, {\tilde N}^Q)$. The asymptotic behavior at
$r\to\infty$ is
\begin{mathletters}
\label{N-l-2}
\begin{eqnarray}
{\tilde N}^M (t,r)
&=&
{\tilde N}_+(t) + O^\infty(r^{-5})
\ \ ,
\\
{\tilde N}^Q (t,r)
&=&
{\tilde \Phi}_+(t)
+ O^\infty(r^{-1})
\ \ ,
\end{eqnarray}
\end{mathletters}%
and the asymptotic behavior at $r=0$ is
\begin{mathletters}
\label{N-s-2}
\begin{eqnarray}
{\tilde N}^M (t,r)
&=&
{\tilde N}^M_0 (t)
+ O(r^2)
\ \ ,
\label{tNM-s-2}
\\
{\tilde N}^Q (t,r)
&=&
{\tilde \Phi}_0(t)
+ O(r^2)
\ \ ,
\end{eqnarray}
\end{mathletters}%
where
\begin{equation}
{\tilde N}^M_0
=
\case{1}{4} N_1 \Lambda_0 R_0^{-1} R_2^{-1} \left( 3 R_0^2 \ell^{-2}
+ 1 - Q_0^2 R_0^{-2} \right)
\ \ .
\end{equation}
When the constraints $M'=0$ and $Q'=0$ hold, Eqs.~(\ref{s2-r-M}),
(\ref{s2-r-Q}), and (\ref{M2}) show that
\begin{equation}
{\tilde N}^M_0
{\buildrel {Q'=0 \atop M'=0} \over =}
N_1 \Lambda_0^{-1}
\ \ .
\label{tNM-s-2-const}
\end{equation}
Thus, when the constraints hold, fixing ${\tilde N}^M$ and ${\tilde
N}^Q$ at $r=0$ is equivalent to fixing $N_1\Lambda_0^{-1}$ and
${\tilde \Phi}_0$. We therefore adopt ${\tilde N}^M$, $N^{\sf R}$,
and ${\tilde N}^Q$ as a set of new independent Lagrange multipliers.

The bulk action takes the form
\begin{eqnarray}
S_\Sigma
[M, {\sf R}, Q, P_M,  P_{\sf R},&& P_Q ; {\tilde N}^M , N^{\sf R},
{\tilde N}^Q]
\nonumber
\\
= \int dt
\int _0^\infty dr
\Biggl\{
&&
P_M { \dot M}
+ P_{\sf R} \dot{\sf R}
+ P_Q {\dot Q}
+ {\tilde N}^Q Q'
-N^{\sf R} P_{\sf R}
\nonumber
\\
&&+ {\tilde N}^M \left[
(1-g)M'
+ 2 g
{\left(3 R_0^2 \ell^{-2} + 1 - Q_0^2 R_0^{-2}\right)}^{-1}
\left( R_0 M' - Q_0 Q' \right)
\right]
\Biggr\}
\ \ .
\nonumber
\\
\label{S2-ham}
\end{eqnarray}
The total action is taken to be
\begin{eqnarray}
S [M, {\sf R}, Q, P_M,  P_{\sf R}, P_Q ; {\tilde N}^M , N^{\sf R},
{\tilde N}^Q]
=
&&
S_\Sigma
[M, {\sf R}, Q, P_M,  P_{\sf R}, P_Q ; {\tilde N}^M , N^{\sf R},
{\tilde N}^Q]
\nonumber
\\
&&
+
S_{\partial\Sigma} [M_0, M_+, Q_0, Q_+ ; {\tilde N}_+, {\tilde
\Phi}_0, {\tilde \Phi}_+]
\ \ ,
\label{S2-total}
\end{eqnarray}
where
\begin{eqnarray}
S_{\partial\Sigma} && [M_0, M_+, Q_0, Q_+ ; {\tilde N}_+, {\tilde
\Phi}_0, {\tilde \Phi}_+]
\nonumber
\\
&&=
\int dt \left(
\casehalf R_0^2 {\tilde N}^M_0
- {\tilde N}_+ M_+
+ {\tilde \Phi}_0 Q_0
- {\tilde \Phi}_+ Q_+ \right)
\ \ .
\label{S2-boundary}
\end{eqnarray}
The quantities to be varied independently are $M$, ${\sf R}$, $Q$,
$P_M$, $P_{\sf R}$, $P_Q$, ${\tilde N}^M$, $N^{\sf R}$, and~${\tilde
N}^Q$, and the boundary conditions for the new Lagrange multipliers are
given by (\ref{N-l-1-R}), (\ref{N-s-1-R}), (\ref{N-l-2}),
and~(\ref{N-s-2}). The volume term in the variation of
(\ref{S2-total}) is proportional to the equations of motion
\begin{mathletters}
\label{eom2}
\begin{eqnarray}
{\dot M} &=& 0
\ \ ,
\\
{\dot {\sf R}} &=& N^{\sf R}
\ \ ,
\\
{\dot Q} &=& 0
\ \ ,
\\
{\dot P}_M &=&
{\left( N^M \right)}'
\ \ ,
\\
{\dot P}_{\sf R} &=& 0
\ \ ,
\\
{\dot P}_Q &=&
{\left( N^Q \right)}'
\ \ ,
\\
M' &=& 0
\ \ ,
\label{eom2-M'}
\\
P_{\sf R} &=& 0
\ \ ,
\label{eom2-PsfR}
\\
Q' &=& 0
\ \ ,
\label{eom2-Q'}
\end{eqnarray}
\end{mathletters}%
where $N^M$ and $N^Q$ are now understood to be defined
by~(\ref{N-def2}). The boundary terms in the variation consist of
terms proportional to $\delta M$, $\delta {\sf R}$, and $\delta Q$ on
the initial and final hypersurfaces, and terms from $r=0$ and
$r=\infty$ given by
\begin{equation}
\int dt \, \left(
\casehalf R_0^2 \,
\delta \! {\tilde N}^M_0
- M_+ \, \delta {\tilde N_+}
+ Q_0 \, \delta {\tilde \Phi}_0
- Q_+ \, \delta {\tilde \Phi}_+
\right)
\ \ .
\label{bt2-0-infty}
\end{equation}
To arrive at~(\ref{bt2-0-infty}), (\ref{deltaR0}) has been used. The
action (\ref{S2-total}) thus yields the equations of motion
(\ref{eom2}) provided that we fix, in addition to the initial and
final values of the new canonical coordinates, also the quantities
${\tilde N}^M_0$, ${\tilde N_+}$, ${\tilde \Phi}_0$, and ${\tilde
\Phi}_+$. Because of~(\ref{tNM-s-2-const}), these fixed quantities at
the right and left ends have precisely the same interpretation
in terms of the geometry of the classical solutions as the fixed
quantities in the action~(\ref{S-total}).

\section{Hamiltonian reduction}
\label{sec:reduction}

In this section we shall reduce the action (\ref{S2-total}) to the
true dynamical degrees of freedom by solving the constraints.

The constraints (\ref{eom2-M'}) and (\ref{eom2-Q'}) imply that $M$
and $Q$ are independent of~$r$. We can therefore write
\begin{mathletters}
\begin{eqnarray}
M(t,r) &=& \bm(t)
\ \ ,
\\
Q(t,r) &=& \bq(t)
\ \ .
\label{bq-def}
\end{eqnarray}
\end{mathletters}%
Substituting this and the constraint (\ref{eom2-PsfR}) back into
(\ref{S2-total}) yields the true Hamiltonian action
\begin{equation}
S [ \bm, \bq, {\bf p}_\bm, {\bf p}_\bq ;
{\tilde N}^M_0 , {\tilde N}_+ , {\tilde \Phi}_+ , {\tilde \Phi}_0 ]
=
\int dt
\left(
{\bf p}_\bm {\dot \bm}
+
{\bf p}_\bq {\dot \bq}
- {\bf h} \right)
\ \ ,
\label{S-red}
\end{equation}
where
\begin{mathletters}
\begin{eqnarray}
{\bf p}_\bm &=& \int_0^\infty dr \, P_M
\ \ ,
\label{bfpm}
\\
{\bf p}_\bq &=& \int_0^\infty dr \, P_Q
\ \ .
\label{bfpq}
\end{eqnarray}
\end{mathletters}%
The reduced Hamiltonian ${\bf h}$ in (\ref{S-red}) is
\begin{equation}
{\bf h} = - \casehalf \Rh^2 {\tilde N}^M_0
+ {\tilde N}_+ \bm +
\left( {\tilde \Phi}_+ - {\tilde \Phi}_0 \right) \bq
\ \ ,
\label{h-red}
\end{equation}
where $\Rh := \Rhor(\bm,\bq)$. The assumptions made in the previous
section imply
\begin{equation}
\bm > M_{\rm crit}(\bq)
\ \ ,
\label{mq-ineq}
\end{equation}
and ${\bf h}$ is therefore well-defined. Note that ${\bf h}$ is,
in general, explicitly time-dependent through the prescribed
functions ${\tilde N}^M_0(t)$, ${\tilde N}_+(t)$, ${\tilde
\Phi}_+(t)$, and~${\tilde \Phi}_0(t)$.

The variational principle associated with the reduced
action~(\ref{S-red}) fixes the initial and final values of the
coordinates $\bm$ and~$\bq$. The equations of motion are
\begin{mathletters}
\label{red-eom}
\begin{eqnarray}
{\dot \bm}
&=& 0
\ \ ,
\label{red-eom-m}
\\
{\dot \bq}
&=& 0
\ \ ,
\label{red-eom-q}
\\
{\dot {\bf p}_\bm}
&=&
2 \Rh
{\left(3 \Rh^2 \ell^{-2} + 1 - \bq^2 R_{\rm
h}^{-2}\right)}^{-1}
{\tilde N}^M_0
- {\tilde N}_+
\ \ ,
\label{red-eom-pm}
\\
{\dot {\bf p}_\bq}
&=&
- 2 \bq
{\left(3 \Rh^2 \ell^{-2} + 1 - \bq^2 R_{\rm
h}^{-2}\right)}^{-1}
{\tilde N}^M_0
+ {\tilde \Phi}_0
- {\tilde \Phi}_+
\ \ .
\label{red-eom-pq}
\end{eqnarray}
\end{mathletters}%
Eqs.\ (\ref{red-eom-m}) and (\ref{red-eom-q}) are readily
understood in terms of the statement that on a classical solution $\bm$
and $\bq$ are respectively equal to the  mass and charge parameters of
the RNAdS solution. To understand Eq.~(\ref{red-eom-pm}), recall from
Sec.\ \ref{sec:transformation} that on a classical solution
$P_M = -T'$, where $T$ is the Killing time. {}From (\ref{bfpm}) we see
that ${\bf p}_\bm = T_0 - T_+$, where $T_0$ and $T_+$ are respectively
the values of $T$ at the left and right ends of the constant $t$
hypersurface. As the constant $t$ hypersurface evolves in the RNAdS
spacetime, the first and second term on the right hand side of
(\ref{red-eom-pm}) are respectively equal to ${\dot T}_0$ and~$-{\dot
T}_+$. The interpretation of Eq.\ (\ref{red-eom-pq}) is analogous. On a
classical solution we have ${\bf p}_\bq = \xi_0 - \xi_+$, where $\xi$ is
the function that specifies the electromagnetic gauge
via~(\ref{Xi-def}). The first two terms on the right hand side of
(\ref{red-eom-pq}) give~${\dot \xi}_0$, and the last term gives~$-{\dot
\xi}_+$.

\section{Quantum theory and the grand partition function}
\label{sec:quantum}

We shall now quantize the reduced Hamiltonian theory of
Sec.~\ref{sec:reduction}\null. Our aim is to construct the
time evolution operator in the Hamiltonian quantum theory, and then to
obtain a grand partition function via an analytic continuation of
this operator.

\subsection{Quantization}
\label{subsec:quantization}

As is well known, the quantization of a given classical Hamiltonian
theory requires input \cite{woodhouse,isham-LH,AAbook}, and the
questions of physically appropriate input for a quantum black hole
remain largely open. For the purposes of the present paper we
shall be content to define the quantum theory in essence by fiat,
following Refs.\ \cite{LW2,kuchar1}. Our main physical conclusions will
emerge from the semiclassical regime of the theory, and at this level
one may reasonably hope the details of the quantization not to be
crucial.

We regard $\bm$ and $\bq$ as configuration variables, satisfying the
inequality~(\ref{mq-ineq}). The wave functions are of the form
$\psi(\bm,\bq)$, and the inner product is taken to be
\begin{equation}
\left( \psi, \chi \right)
= \int_A \mu d\bm d\bq \,
{\overline \psi} \chi
\ \ ,
\label{ip}
\end{equation}
where $A\subset \BbbR^2$ is the domain~(\ref{mq-ineq}) and
$\mu(\bm,\bq)$ is a smooth positive weight factor. The Hilbert space is
thus ${\cal H} := L^2 \bigl( A; \mu d\bm d\bq\bigr)$. We assume
that $\mu$ is a slowly varying function, in a sense to be made more
precise later, but otherwise it will remain arbitrary.

The Hamiltonian operator ${\hat {\bf h}}(t)$ is taken to
act as pointwise multiplication by the function
${\bf h}(\bm,\bq;t)$~(\ref{h-red}):
$\psi(\bm,\bq) \mapsto {\bf h}(\bm,\bq;t) \psi(\bm,\bq)$. ${\hat {\bf
h}}(t)$ is an unbounded essentially self-adjoint
operator \cite{reed-simon}, and the corresponding unitary time evolution
operator in ${\cal H}$ is
\begin{equation}
{\hat K} (t_2; t_1)
=
\exp \left[ -i \int_{t_1}^{t_2} dt' \,
{\hat {\bf h}}(t') \right]
\ \ .
\label{propagator}
\end{equation}
${\hat K} (t_2; t_1)$ acts in ${\cal H}$
by pointwise multiplication by the function
\begin{equation}
K(\bm,\bq;
{\cal T},
\Xi_+, \Xi_0, \Theta )
=
\exp\left[
-i \bm {\cal T}
-i \bq
\left( \Xi_+ - \Xi_0 \right)
+ \casehalf i \Rh^2 \Theta
\right]
\ \ ,
\label{K-function}
\end{equation}
where
\begin{mathletters}
\label{cT-and-Theta}
\begin{eqnarray}
{\cal T}
&:=&
\int_{t_1}^{t_2} dt \, {\tilde N}_+ (t)
\ \ ,
\label{cT-and-Theta:cT}
\\
\Xi_+
&:=&
\int_{t_1}^{t_2} dt \, {\tilde \Phi}_+ (t)
\ \ ,
\\
\Xi_0
&:=&
\int_{t_1}^{t_2} dt \, {\tilde \Phi}_0 (t)
\ \ ,
\\
\Theta
&:=&
\int_{t_1}^{t_2} dt \, {\tilde N}^M_0 (t)
\ \ .
\end{eqnarray}
\end{mathletters}%
${\hat K} (t_2; t_1)$ therefore depends on $t_1$ and $t_2$ only
through the quantities on the left hand side of~(\ref{cT-and-Theta}),
and we may write ${\hat K} (t_2; t_1)$ as ${\hat K} ({\cal T}, \Xi_+,
\Xi_0, \Theta)$. The composition law,
${\hat K} (t_3; t_2) {\hat K} (t_2; t_1)
= {\hat K} (t_3; t_1)$,
amounts to independent addition in each of the four
parameters in ${\hat K} ({\cal T}, \Xi_+, \Xi_0, \Theta)$, and we may
regard these four parameters as independent evolution parameters
specified by the boundary conditions. ${\cal T}$~is the Killing time
elapsed at the infinity, and $\Theta$ is the boost parameter elapsed
at the bifurcation two-sphere. $\Xi_+$~and $\Xi_0$ can be computed from
the line integral of the electromagnetic potential (\ref{4-vectorpot})
along the timelike curve of constant $r$ and constant angular variables
as this curve approaches respectively the infinity and the bifurcation
two-sphere.

\subsection{Grand partition function}
\label{subsec:grand-partition}

We shall now construct a grand partition function by continuing the time
evolution operator to imaginary time and taking the trace. We begin by
discussing the boundary conditions for the relevant thermodynamical
ensemble.

The envisaged semiclassical thermodynamical situation consists of a
charged spherically symmetric black hole in asymptotically \ads\ space,
in thermal equilibrium with a bath of Hawking radiation. If the
back-reaction from the radiation is neglected, the geometry is
described by the RNAdS metric~(\ref{rnads-metric}). Assuming that
the local temperature is given in the usual manner in terms of the
surface gravity and the redshift factor \cite{pagerev,HPads}, we see
that the local temperature is
$F^{-1/2}\beta^{-1}$, where $F$ is given by~(\ref{rnads-F})
and
\begin{equation}
\beta :=
4\pi R_0
{\left( 3 \ell^{-2} R_0^2 + 1 - Q^2 R_0^{-2} \right)}^{-1}
\ \ .
\label{beta-infty}
\end{equation}
At the infinity the local temperature vanishes as $\beta^{-1} \ell
R^{-1}\left(1 + O^\infty (\ell^2 R^{-2}) \right)$, and $\beta^{-1}$ can
thus be extracted from the asymptotic behavior as the coefficient of
the leading order term~$\ell R^{-1}$. We shall follow
Refs.\ \cite{pagerev,HPads} and regard $\beta^{-1}$ as a renormalized
temperature at infinity.

The electromagnetic variable with thermodynamic interest for us is the
electric potential difference between the horizon and the infinity, in
the curvature coordinates (\ref{rnads-metric}) and in an
electromagnetic gauge that makes $A$ invariant under the Killing time
translations. We denote this quantity by~$\phi$. {}From (\ref{rnads-A})
it is seen that on a classical solution $\phi=Q R_0^{-1}$.

We shall consider a thermodynamical ensemble in which the fixed
quantities are $\beta$ and~$\phi$. This data can be
interpreted as that for a grand canonical ensemble, with $\phi$
being analogous to the chemical potential \cite{davies1,davies2,BBWY}.
Our aim is to obtain a grand partition function ${\cal
Z}(\beta,\phi)$ by continuing the time evolution operator
of the Lorentzian Hamiltonian theory to imaginary time and taking the
trace.

The continuation of ${\cal T}$ is straightforward: comparing the
definition of $\beta$ to the falloff of $N$ in (\ref{l-r}) and
to the definition~(\ref{cT-and-Theta:cT}), we are led to set
${\cal T}=-i\beta$. For the continuation of $\Theta$ we choose
$\Theta=-2\pi i$, motivated by the regularity of the classical
Euclidean solutions as in Ref.\ \cite{LW2}. We mentioned at the end of
Sec.\ \ref{sec:metric} that the regularity of the electromagnetic
potential at the bifurcation two-sphere of the Lorentzian solutions
requires ${\tilde \Phi}_0=0$; similarly, requiring regularity of the
electromagnetic potential at the horizon of the classical Euclidean
solutions now leads us to set $\Xi_0=0$. Finally, recall that $\Xi_+$
gives the constant $r$ line integral of the electromagnetic potential
(\ref{4-vectorpot}) at the infinity. Comparing this to the definition
of~$\phi$, we set $\Xi_+=- {\cal T} \phi = i\beta \phi$. We are thus
led to propose for the grand partition function the expression
\begin{equation}
{\cal Z}(\beta,\phi)
=
{\rm Tr} \left[ {\hat K} (-i\beta ,
i\beta \phi ,
0 ,
-2\pi i) \right]
\ \ .
\label{cZ-trace}
\end{equation}
As it stands, the trace in (\ref{cZ-trace}) is divergent, but one can
argue as in Refs.\ \cite{LW2,BLPP} that a suitable regularization and
renormalization yields the result
\begin{equation}
\czren (\beta,\phi)
=
{\cal N} \int_A \mu d\bm d\bq \,
\exp\left[
-\beta \left( \bm - \bq \phi \right)
+ \pi \Rh^2 \right]
\ \ ,
\label{cZ-ren}
\end{equation}
where we have substituted for $K$ the explicit
expression~(\ref{K-function}). The normalization factor ${\cal N}$ may
depend on~$\ell$, but we shall assume that it does not depend on $\beta$
or~$\phi$.

Provided the weight factor $\mu$ is slowly varying compared with the
exponential in~(\ref{cZ-ren}), it is easy to verify, using the
definition of $\Rh$ given after Eq.~(\ref{h-red}), that the integral in
(\ref{cZ-ren}) is convergent. Equation (\ref{cZ-ren}) thus yields a
well-defined grand partition function. Comparing with ordinary PVT
systems \cite{reichl}, $\phi$ is now indeed seen to be analogous to the
chemical potential, and the quantities $\bm$ and $\bq$ are respectively
analogous to the energy and the particle number. We shall examine the
thermodynamical properties of this grand partition function in the next
section.

\section{Thermodynamics in the grand canonical ensemble}
\label{sec:thermo-grand}

It is useful to change the integration variables in (\ref{cZ-ren})
from the pair $(\bm,\bq)$ to the pair $(\Rh,\bq)$. {}From
(\ref{massfunc}) we obtain
\begin{equation}
\bm =
\casehalf \Rh
\left( \Rh^2 \ell^{-2} + 1 + \bq^2 \Rh^{-2} \right)
\ \ ,
\label{massfunc2}
\end{equation}
and the grand partition function takes the form
\begin{equation}
\czren (\beta,\phi)
= {\cal N} \int_{A'} {\tilde \mu}
\, d\Rh d\bq \,
\exp\left(  -I_* \right)
\ \ ,
\label{cZ-istar}
\end{equation}
where
\begin{equation}
I_* (\Rh,\bq) :=
\casehalf \beta \Rh
\left( \Rh^2 \ell^{-2} + 1 + \bq^2 \Rh^{-2} \right)
- \beta \phi \bq - \pi \Rh^2
\ \ .
\label{istar}
\end{equation}
One may view $I_*$ as an effective action or a reduced
action \cite{WYprl,whitingCQG,BBWY}.
The integration domain $A'$ is given
by the inequalities
\begin{mathletters}
\label{Aprime}
\begin{eqnarray}
&&0 \le \Rh
\ \ ,
\\
&&\bq^2 \le \Rh^2 \left( 1 + 3 \Rh^2 \ell^{-2} \right)
\ \ ,
\label{ineqb}
\end{eqnarray}
\end{mathletters}%
and the weight factor $\tilde\mu$ is obtained from $\mu$ by including
the Jacobian $\left| \partial(\bm,\bq) / \partial(\Rh,\bq) \right|$.
Note that because of~(\ref{ineqb}), $I_*$~remains finite as $\Rh\to0$.

As $\tilde\mu$ is assumed to be slowly varying, we can estimate $\czren
(\beta,\phi)$  by the saddle point approximation to~(\ref{cZ-istar}).
For this, we need to find the critical points of $I_*$ in the
interior of~$A'$.

When $\phi^2 < 1 - \case{4}{3}\pi^2 \ell^2 \beta^{-2}$, $I_*$~has no
critical points. When $1 - \case{4}{3}\pi^2 \ell^2 \beta^{-2} < \phi^2
< 1$, the two critical points of $I_*$ are at
\begin{mathletters}
\label{critical}
\begin{eqnarray}
\Rh = \Rh^\pm &:=& {2\pi \ell^2 \over 3 \beta}
\left( 1 \pm \sqrt{1 +  {3\beta^2 (\phi^2-1) \over 4 \pi^2 \ell^2}}
\right)
\ \ ,
\\
\bq = \bq^\pm &:=& \phi \Rh^\pm
\ \ .
\end{eqnarray}
\end{mathletters}%
The lower signs do not give a local extremum, but the upper signs give
a local minimum. In the limiting case $1 - \case{4}{3}\pi^2 \ell^2
\beta^{-2} = \phi^2$, the only critical point is $(\Rh^+,\bq^+)$, but
it is not a local extremum. Finally, when $\phi^2\ge1$, the only
critical point is $(\Rh^+,\bq^+)$, and it is a local minimum. Whenever
the critical points exist, the value of $I_*$ at these points can be
written as
\begin{equation}
I_*(\Rh^\pm,\bq^\pm) =
{ \pi {(\Rh^\pm)}^2 \left( 1 - \phi^2 - {(\Rh^\pm)}^2 \ell^{-2}
\right)
\over
1 - \phi^2 + 3{(\Rh^\pm)}^2 \ell^{-2}
}
\ \ .
\end{equation}

As $I_*$ grows without bound in the noncompact directions in~$A'$, the
global minimum can be found by examining $I_*$ at the critical points
and on the boundary of~$A'$. When $\phi^2 > 1 -
\pi^2\ell^2\beta^{-2}$, the global minimum  is at the critical point
$(\Rh^+,\bq^+)$, and $I_*(\Rh^+,\bq^+)$ is negative. When
$\phi^2 < 1 - \pi^2 \ell^2\beta^{-2}$, the global minimum is at $\Rh =
0 = \bq$, where $I_*$ vanishes. In the limiting case $\phi^2 = 1 -
\pi^2 \ell^2\beta^{-2}$, $I_*$ vanishes at $(\Rh^+,\bq^+)$ and at $\Rh
= 0 = \bq$, and is positive everywhere else.

We thus see that for $\phi^2 > 1 - \pi^2\ell^2\beta^{-2}$, $\czren$
can be approximated as
\begin{equation}
\czren(\beta,\phi) \approx P \exp[ -I_*(\Rh^+,\bq^+) ]
\ \ ,
\label{czrenapp}
\end{equation}
where $P$ is a slowly varying prefactor. The approximation becomes
presumably progressively better with increasing $|I_*(\Rh^+,\bq^+)|$.
For $\phi^2 < 1 - \pi^2\ell^2\beta^{-2}$, the dominant
contribution to $\czren$ comes from the vicinity of $\Rh = 0 = \bq$, and
the behavior of $\czren$ depends more sensitively on the weight
factor~$\tilde\mu$.

These results for $\czren$ are consistent with what one would
expect just from the existence of (Lorentzian) black hole solutions
under fixing $\phi$ and the renormalized inverse Hawking temperature
$\beta$~(\ref{beta-infty}). It can be verified that such solutions
exist precisely at the critical points of~$I_*$: the values of $\bm$
and $\bq$ at these critical points are just the mass and charge
parameters of the black hole. Further, the value of $I_*$ at a critical
point is simply the Euclidean action of the corresponding Euclideanized
black hole solution. When a unique classical solution exists, it
dominates the grand partition function; when two distinct classical
solutions exist, the grand partition function is dominated either by
the larger mass classical solution or by no classical solutions. The
situation is thus remarkably similar to that found in the absence of a
cosmological constant when the boundary conditions are set on a finite
size box \cite{BBWY}.

Let us now consider the thermodynamical predictions from~$\czren$.
Recall that the thermal expectation values of the energy and charge in
the grand canonical ensemble are given by
\begin{mathletters}
\begin{eqnarray}
\langle E \rangle &=&
\left( - {\partial \over \partial \beta}
+ \beta^{-1} \phi {\partial \over \partial \phi} \right)
(\ln \czren)
\ \ ,
\\
\langle Q \rangle &=&
\beta^{-1} {\partial (\ln \czren) \over \partial \phi}
\ \ .
\label{Qexact}
\end{eqnarray}
\end{mathletters}%
When $\czren$ is dominated by the critical point $(\Rh^+,\bq^+)$, we
find
\begin{mathletters}
\label{EQapprox}
\begin{eqnarray}
\langle E \rangle &\approx& \bm^+
\ \ ,
\label{Eapprox}
\\
\langle Q \rangle &\approx&
\bq^+
\ \ ,
\label{Qapprox}
\end{eqnarray}
\end{mathletters}%
where $\bm^+$ is obtained from $(\Rh^+,\bq^+)$
through~(\ref{massfunc2}). That is, the thermal expectation values of
the energy and the charge are simply the mass and charge parameters of
the dominant classical solution. In particular, there are no additional
contributions to the mass from the gravitational binding energy
associated with the thermal energy, or from the electrostatic binding
energy associated with the charge. Such additional, finite size
contributions were found to be present in the finite size ensembles of
Refs.\ \cite{york1,WYprl,BBWY,BLPP}.

It is easily seen that $(\partial \bm^+/\partial \beta)<0$. This means
that when the approximation (\ref{Eapprox}) is good, the (constant
$\phi$) heat capacity, $C_\phi = - \beta^2 (\partial \langle E \rangle
/ \partial\beta)$, is positive. In the regime~(\ref{Eapprox}), the
system is thus stable under thermal fluctuations in the energy. Note
that as $(\partial \bm^-/\partial \beta)>0$, a grand partition function
dominated by the lower mass classical solution would be
thermodynamically unstable \cite{whitingCQG}. This is analogous to what
happens in the absence of a cosmological constant under the boxed
boundary conditions considered in Refs.\ \cite{york1,WYprl,BBWY}.

It is also easily seen that $(\partial \bq^+/\partial\phi)>0$. This
shows that when the approximation (\ref{Qapprox}) is good, we have
$(\partial\langle Q \rangle / \partial \phi)>0$, and the system is
stable under thermal fluctuations in the charge. More generally, one
can show directly from the expressions (\ref{cZ-istar}), (\ref{istar}),
and (\ref{Qexact}) that $(\partial\langle Q \rangle /
\partial \phi)>0$ holds always, even when the approximation
(\ref{Qapprox}) is not good.

The entropy in the grand canonical ensemble is given by
\begin{equation}
S = \left( 1 - \beta {\partial \over \partial \beta} \right)
(\ln \czren)
\ \ .
\end{equation}
When the approximation (\ref{czrenapp}) is good, we have $S \approx
\pi {(\Rh^+)}^2$, which means that the entropy is one quarter of the
horizon area. This is the anticipated Bekenstein-Hawking result.

Finally, when $\czren$ is not dominated by a critical point, the
thermodynamical predictions become much more sensitive to the choice
of the weight factor~$\tilde\mu$. As in
Refs.\ \cite{WYprl,whitingCQG,BBWY,BLPP}, one can view the transition
in the qualitative behavior of $\czren$ as evidence for a phase
transition between a black hole sector and a topologically different
sector of the theory; in the case at hand, the second sector might be
referred to as ``hot \ads\ space."  On classical grounds one might have
expected this transition to occur near $\phi^2 \approx 1 -
\case{4}{3}\pi^2 \ell^2 \beta^{-2}$, where the classical solutions
disappear. However, we saw that the transition in fact occurs near
$\phi^2 \approx 1 - \pi^2 \ell^2 \beta^{-2}$, where the two classical
solutions still exist. This is highly similar to what happens in four
dimensions under boxed boundary conditions without a cosmological
constant \cite{WYprl,BBWY},
but subtly different from what happens in two
dimensions with Witten's dilatonic black hole \cite{BLPP}.

\section{The canonical ensemble}
\label{sec:can-ensemble}

We have seen that the Hamiltonian formulation of Secs.\
\ref{sec:metric}--\ref{sec:reduction} led into a thermodynamical grand
canonical ensemble where the fixed quantities are the renormalized
inverse temperature $\beta$ at infinity and the electric potential
difference $\phi$ between the horizon and the infinity. {}From the
thermodynamical viewpoint, another natural ensemble for the charged
black hole in asymptotically \ads\ space is the canonical ensemble,
where one allows fluctuations in $\phi$ but fixes instead the
charge~$\bq$. In this section we shall outline the recovery
of the canonical ensemble from a Lorentzian Hamiltonian analysis, and
briefly discuss the thermodynamical properties of the black hole
in this ensemble.

As a starting point, we modify the boundary conditions of
the Hamiltonian theory of Sec.\ \ref{sec:metric} by leaving
${\tilde\Phi}_0(t)$ and ${\tilde\Phi}_+(t)$ unspecified but fixing
$Q_0(t)$ and $Q_+(t)$ to be prescribed functions of~$t$. The action is
obtained from (\ref{S-total}) and (\ref{S-boundary}) by omitting the
terms $\int dt \left( {\tilde \Phi}_0 Q_0 - {\tilde \Phi}_+ Q_+
\right)$ from~(\ref{S-boundary}). Clearly, classical solutions exist
only when $Q_0(t)$ and $Q_+(t)$ are chosen independent of $t$ and
equal. We shall from now on assume that the boundary data is chosen in
this manner.

One way to proceed is simply to push through the canonical
transformation of Sec.~\ref{sec:transformation}, noting that the
new boundary conditions merely result into minor modifications. It is
only when one subsequently performs a Hamiltonian reduction along
the lines of Sec.~\ref{sec:reduction} that the new boundary conditions
give rise to important differences. Firstly, the boundary data for
$Q_0(t)$ and $Q_+(t)$ implies that the quantity $\bq(t)$ defined by
(\ref{bq-def}) is a $t$-independent constant whose value is completely
determined by the boundary conditions. Therefore, the Liouville term
$\int dt \, {\bf p}_\bq {\dot \bq}$ drops entirely out of the
action~(\ref{S-red}). Secondly, because of the terms that were omitted
from the boundary action~(\ref{S-boundary}), the term $\left(
{\tilde\Phi}_+ - {\tilde\Phi}_0 \right) \bq$ drops out of the reduced
Hamiltonian~(\ref{h-red}). This means that in the reduced Hamiltonian
theory $\bq$ has become an external parameter specified by the boundary
conditions: it is not varied in the action, and it does not have a
conjugate momentum. The new reduced action reads
\begin{equation}
S_C [ \bm, {\bf p}_\bm ; {\tilde N}^M_0 , {\tilde N}_+ ; \bq ]
=
\int dt
\left( {\bf p}_\bm {\dot \bm} - {\bf h}_C \right)
\ \ ,
\label{SC-red}
\end{equation}
where
\begin{equation}
{\bf h}_C = - \casehalf \Rh^2 {\tilde N}^M_0  +
{\tilde N}_+ \bm
\ \ .
\label{hC-red}
\end{equation}
Here $\Rh := \Rhor(\bm,\bq)$ as before, and the assumptions made in
the canonical transformation again imply that (\ref{mq-ineq}) holds.

An alternative way to proceed under the new boundary data is to
partially reduce the action already in the variables of Sec.\
\ref{sec:metric} by solving the constraint~(\ref{constr-eqs-G}). One
uses the constraint (\ref{constr-eqs-G}) and the equation of motion
(\ref{dyn-eqs-PB}) to set $P_\Gamma(t,r)$ equal to the constant
specified in the boundary data, and substitutes this back in the
action. The Liouville term $\int_0^\infty dr \, P_\Gamma {\dot
\Gamma}$ then becomes a total time derivative and can be dropped. One
thus obtains an action that no longer involves $\Gamma$
or~${\tilde\Phi}$, involves $P_\Gamma$ only as a prescribed constant,
and correctly yields the equations of motion for the remaining
variables. One can now perform a canonical transformation from the
variables
$(\Lambda,R,P_\Lambda,P_R)$ to the new variables
$(M,{\sf R},P_M,P_{\sf R})$, defined as in Sec.\
\ref{sec:transformation} except that $P_\Gamma=Q$ is now regarded as a
fixed external parameter. Finally, one can reduce the action by
solving the constraints as in Sec.~\ref{sec:reduction}\null. The result
is again the action given by (\ref{SC-red}) and~(\ref{hC-red}).

Quantization of the reduced Hamiltonian theory proceeds as in
Sec.~\ref{sec:quantum}\null. For the renormalized trace of the
analytically continued time evolution operator, we obtain
\begin{equation}
\zren (\beta,\bq) = \int\limits_{\Rcrit(\bq)}^\infty
{\tilde{\tilde\mu}}
\, d\Rh \,
\exp\left(  -I_{C*} \right)
\ \ ,
\label{Z-istarc}
\end{equation}
where the function $\Rcrit$ is defined by Eq.~(\ref{Rcrit}) in
Appendix~\ref{app:rnads}, the weight factor ${\tilde{\tilde\mu}}$
is a positive function of $\Rh$ (and possibly~$\bq$), and
\begin{equation}
I_{C*} (\Rh) :=
\casehalf \beta \Rh
\left( \Rh^2 \ell^{-2} + 1 + \bq^2 \Rh^{-2} \right)
- \pi \Rh^2
\ \ .
\label{istarc}
\end{equation}

Under the assumption that ${\tilde{\tilde\mu}}$ is slowly varying,
the dominant contribution to $\zren$ can be estimated by saddle point
methods. The cases $\bq=0$ and $\bq\ne0$ merit each a separate
analysis.

Consider first the special case $\bq=0$. The lower limit of the integral
in (\ref{Z-istarc}) is then at $\Rh=0$. The critical point structure of
$I_{C*}$ is identical to that of $I_{*}$ (\ref{istarc})
for $\phi=0$, and the locations of the critical points and the values of
the action at these points can simply be read off from Sec.\
\ref{sec:thermo-grand} by setting $\phi=0$.

Consider from now on the generic case $\bq\ne0$. $I_{C*}$ has one
negative critical point, and from one to three positive critical
points. The negative critical point is unphysical, but all the positive
critical points lie in the physical domain $\Rh>\Rcrit(\bq)$. As
$I_{C*}$ is decreasing at $\Rh=\Rcrit(\bq)$ and tends to infinity as
$\Rh\to\infty$, the global minimum of $I_{C*}$ in the domain
$\Rh>\Rcrit(\bq)$ is at a critical point. We can therefore
concentrate on the positive critical points.

When $\beta^2 \ge \case{3}{2}\pi^2 \ell^2$, $I_{C*}$ has only one
positive critical point. When $\beta^2 < \case{3}{2}\pi^2 \ell^2$, the
number of positive critical points is determined by the status of the
double inequality
\begin{equation}
{(1-3s)(1+s) \over 36 {(1-s)}^2}
\le \bq^2 \ell^{-2} \le
{(1+3s)(1-s) \over 36 {(1+s)}^2}
\ \ ,
\label{s-ineq}
\end{equation}
where
\begin{equation}
s := \sqrt{1 - {2\beta^2 \over 3 \pi^2 \ell^2}}
\ \ .
\end{equation}
When (\ref{s-ineq}) does not hold, there is only one positive critical
point. When (\ref{s-ineq}) holds as a genuine inequality, there are
three positive critical points, and saturating the inequalities gives
limiting cases where two of the three positive critical points merge.
(Note that if $\beta^2 \le \case{4}{3}\pi^2 \ell^2$, the leftmost
expression in (\ref{s-ineq}) is non-positive, and the left hand side
inequality is then necessarily genuinely satisfied.) Now, when only
one positive critical point exists, this critical point is the global
minimum. On the other hand, when three positive critical points exist,
they constitute a local maximum between two local minima, and the
global minimum can be at either of the local minima depending on the
values of the parameters. For example, when the right hand side
inequality in (\ref{s-ineq}) is close to being saturated, the global
minimum is at the local minimum with the larger value of~$\Rh$.

The critical points can be examined further by parametrizing $\beta$ and
$\bq$ as
\begin{mathletters}
\label{uv-para}
\begin{eqnarray}
4\pi \ell \beta^{-1}
&=&
{(u-v) \left[ 3 \left( u^2 + v^2 \right) + 1 \right]
\over
u^2 + v^2 - uv}
\ \ ,
\\
\noalign{\smallskip}
\bq^2 \ell^{-2}
&=&
{u^2 v^2 (3uv+1)
\over
u^2 + v^2 - uv}
\ \ ,
\end{eqnarray}
\end{mathletters}%
where the parameters $u$ and $v$ satisfy $0<v<u$. The negative,
unphysical critical point is then at $\Rh=-\ell v$, and $\Rh=\ell u$
gives a positive critical point.\footnote{We thank Bernard Whiting for
suggesting this type of parametrization.}
The condition that only one positive critical point exist reads
\begin{equation}
\alpha(u) < v
\ \ ,
\label{uv-onecrit}
\end{equation}
where $\alpha(u)$ is the unique solution to the equation
\begin{equation}
0 =
9u \alpha^3 - \left( 6u^2 + 1 \right) \alpha^2
+ u \left( 9u^2 + 2 \right) \alpha - u^2
\end{equation}
in the interval $0<\alpha<u$. In this case the parametrization
(\ref{uv-para}) is unique. When the inequality in (\ref{uv-onecrit}) is
reversed and three positive critical points exist, the parametrization
(\ref{uv-para}) can be made unique by imposing the conditions
\begin{mathletters}
\begin{eqnarray}
u &<& {1 \over \sqrt{3}}
\ \ ,
\\
\noalign{\smallskip}
v &<&
{ \sqrt{
\left( 1 + 6u^2 \right)
\left( 1 - 3u^2 \right) }
- \left( 1 - 3u^2 \right)
\over
9u}
\ \ ,
\end{eqnarray}
\end{mathletters}%
which make $\Rh=\ell u$ the local maximum. The two local minima are
then at the roots of the quadratic equation
\begin{equation}
0=
3\left( u^2 + v^2 - uv \right)
{\left( \Rh/\ell \right)}^2
- (u-v) (3uv+1)
\left( \Rh/\ell \right)
+ uv (3uv+1)
\ \ .
\end{equation}
The global minimum is at the larger (smaller) local
minimum when the inequality
\begin{equation}
0<
12 (6uv-1)
{\left( u^2 + v^2 - uv \right)}^2
+ {(u-v)}^2 (3uv+1) \left[ 3\left( u^2+v^2 \right) + 1 \right]
\end{equation}
is satisfied (reversed).

It is of some interest to examine the behavior of the critical points
in the limit $\bq^2\to0$ with fixed~$\beta$. When $\beta^2 >
\case{4}{3}\pi^2 \ell^2$, the above discussion shows that for
sufficiently small $\bq^2$ there exists only one positive critical
point, and in the limit $\bq^2\to0$ this critical point approaches
zero as
\begin{equation}
\Rh = |\bq| \left[ 1 + 2\pi \beta^{-1} |\bq|
+ O \left( \bq^2
\ell^{-2} \right) \right]
\ \ .
\label{small-q}
\end{equation}
When $\beta^2 < \case{4}{3}\pi^2 \ell^2$, on the other hand, there
exist three positive critical points for sufficiently small~$\bq^2$.
In the limit $\bq^2\to0$, the smallest positive critical point again
approaches zero as~(\ref{small-q}), whereas the two larger ones approach
the two critical points of the case $\bq=0$. In the limiting case
$\beta^2 = \case{4}{3}\pi^2 \ell^2$, the smallest of the three positive
critical points once again approaches zero as~(\ref{small-q}), and the
two larger ones merge into a $\bq=0$ critical point that is not a local
extremum. The limiting behavior is thus smooth, in spite of the
changing number of critical points.

At any critical point, the value of the action can be written as
\begin{equation}
I_{C*}^c =
- {
\pi \Rh^2 \left( \Rh^2 \ell^{-2} -1 - 3 \bq^2 \Rh^{-2} \right)
\over
3 \Rh^2 \ell^{-2} +1 -  \bq^2 \Rh^{-2}
}
\ \ .
\end{equation}
In the limit $\bq\to0$, this agrees with the expression given in
Ref.\ \cite{HPads}.

We thus see that for generic values of the parameters, $\zren
(\beta,\bq)$ can be approximated by $\exp\left[-I_{C*}^{\rm
min}\right]$, where $I_{C*}^{\rm min}$ stands for the value of $I_{C*}$
at the critical point that is the global minimum. As in
Sec.~\ref{sec:thermo-grand}, this is consistent with what one would
have expected just from the existence of (Lorentzian) black hole
solutions under fixing the charge and the renormalized inverse Hawking
temperature: such solutions exist precisely at the critical points
of~$I_{C*}$, and the values of $\bm$ and $\bq$ at these critical points
are just the mass and charge parameters of the black hole. One may view
the shifting of the global minimum of $I_{C*}$ from one local minimum
to the other as a thermodynamical phase transition.

We end this section with some brief remarks on the thermodynamics in the
canonical ensemble. Recall that the formulas for the thermal
expectation values for the energy and the electric potential read
\begin{mathletters}
\begin{eqnarray}
\langle E \rangle &=&
 - {\partial (\ln \zren) \over \partial \beta}
\ \ ,
\\
\langle \phi \rangle &=&
- \beta^{-1} {\partial (\ln \zren) \over \partial \bq}
\ \ .
\label{phi-exact}
\end{eqnarray}
\end{mathletters}%
When a critical point of $I_{C*}$ dominates, we obtain
\begin{mathletters}
\label{Ephiapprox}
\begin{eqnarray}
\langle E \rangle
&\approx&
\bm
\ \ ,
\label{mass-can-approx}
\\
\langle \phi \rangle
&\approx&
{\bq \over
\Rh}
\ \ .
\end{eqnarray}
\end{mathletters}%
These are, respectively, just the mass and the electric potential
difference between the horizon and the infinity for the dominating
classical solution. When the approximation (\ref{mass-can-approx}) is
good, the positivity of the (constant $\bq$) heat capacity, $C_\bq = -
\beta^2 (\partial
\langle E \rangle / \partial\beta)$, follows from the fact that the
dominant critical point is a minimum of
$I_{C*}$ \cite{WYprl,whitingCQG}.
The positivity of $C_\bq$ follows more
generally, even when the saddle point approximation does not hold, by
direct manipulations from the expression (\ref{Z-istarc}) and the
assumption that ${\tilde{\tilde\mu}}$ is positive.

When the saddle point approximation is good, we have for the entropy
the Bekenstein-Hawking result, $S = \left( 1 - \beta (\partial /
\partial\beta)\right) (\ln \zren) \approx \pi {(\Rh)}^2$.

\section{Conclusions and discussion}
\label{sec:discussion}

In this paper we have investigated the Hamiltonian dynamics and
thermodynamics of spherically symmetric Einstein-Maxwell theory with a
negative cosmological constant. We first set up a classical Lorentzian
Hamiltonian theory in which the right end of the spacelike
hypersurfaces is at the asymptotically \ads\ infinity in an exterior
region of a RNAdS black hole spacetime, and the left end of the
hypersurfaces is at the bifurcation two-sphere of a nondegenerate
Killing horizon. We then simplified the constraints by a canonical
transformation, and we explicitly reduced the theory into an
unconstrained Hamiltonian theory with two canonical pairs of degrees of
freedom. The reduced theory was quantized by Hamiltonian methods, and a
grand partition function for a thermodynamical grand canonical ensemble
was obtained by analytically continuing the Schr\"odinger picture time
evolution operator to imaginary time and taking the trace. The analytic
continuation at the bifurcation two-sphere was done in a way motivated
by the smoothness of Euclidean black hole geometries as in
Ref.\ \cite{LW2}. A similar analysis with minor modifications to the
boundary conditions led to a partition function for a thermodynamical
canonical ensemble. Both the canonical ensemble and the grand canonical
ensemble turned out to be well defined, and we were able to find the
conditions under which the (grand) partition function is dominated by a
classical Euclidean solution.

Both thermodynamical ensembles exhibited a phase transition. In the
grand canonical ensemble the transition occurs when the grand
partition function ceases to be dominated by any classical Euclidean
black hole solution, in close analogy with what happens in the
spherically symmetric vacuum canonical ensemble with a finite
boundary \cite{york1,WYprl,whitingCQG}. In the canonical ensemble this
kind of a phase transition can occur only in the limit of a vanishing
charge, whereas for nonvanishing charge there occurs a phase transition
in which the dominating contribution to the partition function shifts
from one classical Euclidean solution to another as the boundary data
changes. In either ensemble, whenever the (grand) partition function is
dominated by a classical solution, one recovers for the entropy the
Bekenstein-Hawking value of one quarter of the horizon area.

The classical canonical transformation of Sec.\ \ref{sec:transformation}
is a relatively straightforward generalization of the
transformation that was found by Kucha\v{r} in the spherically symmetric
vacuum Einstein theory under Kruskal-like boundary
conditions \cite{kuchar1}. When the classical equations of
motion hold, our new canonical coordinates $M$ and $Q$ are simply the
mass and charge parameters of the RNAdS solution. By (generalized)
Birkhoff's theorem, the spacetime is uniquely characterized by these two
parameters and the cosmological constant. The conjugate momenta,
$P_M$ and~$P_Q$, carry the information about the embedding of the
spacelike hypersurface in the spacetime and the electromagnetic gauge.
Upon elimination of the constraints, we saw in Sec.\
\ref{sec:reduction} that $P_M$ and $P_Q$ each give rise to one
unconstrained momentum in the reduced Hamiltonian theory. These reduced
momenta are global constructs with no local geometrical meaning, and
they are associated with the anchoring of the spacelike hypersurfaces
at the infinity and at the bifurcation two-sphere. The electromagnetic
pair $(Q,P_Q)$ is quite closely analogous to the gravitational pair
$(M,P_M)$. The third canonical pair, $({\sf R},P_{\sf R})$, is entirely
gauge, and it completely disappears when the constraints are
eliminated.

Although we have here discussed the canonical transformation only
under boundary conditions motivated by our thermodynamical goal, it
would appear possible to use arguments similar to those in
Refs.\ \cite{kuchar1,varadarajan}
to adapt this canonical transformation
to boundary conditions under which the spacelike hypersurfaces
extend from a left hand side asymptotically \ads\ region to a right
hand side asymptotically \ads\ region, crossing the event horizons in
arbitrary ways. The form (\ref{constr-trans1}) taken by the constraints
then suggests that, after introducing electromagnetic variables
analogous to the reparametrization clocks $\tau_\pm$ of
Ref.\ \cite{kuchar1}, it is possible to perform a canonical
transformation that separates $Q$ into the charge density $Q'$ and the
charge at the (say) left hand side infinity, in analogy with the
transformation that in Ref.\ \cite{kuchar1} separates $M$ into the mass
density $M'$ and  the mass at the left hand side infinity. Also, it
appears possible to take the limit where the cosmological constant
vanishes and the asymptotically \ads\ regions are replaced by
asymptotically flat regions.\footnote{Before the work reported in this
paper was begun, we were informed by Karel Kucha\v{r} that he had
generalized the canonical transformation of Ref.\ \cite{kuchar1} to
the spherically symmetric Einstein-Maxwell system without a
cosmological constant \cite{kuchar-private}. We thank Karel Kucha\v{r}
for correspondence on this point.}  It would further be possible to
consider boundary conditions of the kind put forward in
Ref.\ \cite{marolf-boundary}.  We have not investigated these issues in a
systematic fashion; however, we shall outline in Appendix
\ref{app:as-flat} how our canonical transformation can be adapted to
the limit of a vanishing cosmological constant, under boundary
conditions that still keep the left end of the hypersurfaces at the
bifurcation two-sphere of a nondegenerate Killing horizon but replace
the asymptotically \ads\ falloff conditions at the right end by
asymptotically flat falloff conditions. In this case, each classical
solution is the exterior region of a non-extremal Reissner-Nordstr\"om
black hole.

The thermodynamical results of Secs.\ \ref{sec:thermo-grand} and
\ref{sec:can-ensemble} show that the stabilizing effect of the
negative cosmological constant is highly similar to the stabilizing
effect of a finite ``box" with fixed surface area and fixed local
temperature \cite{york1,WYprl,whitingCQG,BBWY,BLPP}. One important
difference is, however, that in the asymptotically \ads\ case various
thermal expectation values are more directly related to the parameters
of the dominant classical solutions. In the grand canonical ensemble,
equations (\ref{EQapprox}) show that the thermal expectation values of
energy and charge are simply the mass and charge parameters of the
dominant classical solution: there are no additional contributions to
the mass from the gravitational binding energy associated with the
thermal energy, or from the electrostatic binding energy associated
with the charge. Such additional, finite size contributions were found
to be present in the finite size ensembles considered in
Refs.\ \cite{york1,WYprl,BBWY,BLPP}. In the canonical ensemble, the
situation is similar with the thermal expectation values of the energy
and the electric potential~(\ref{Ephiapprox}).

The stabilizing effect of the negative cosmological constant becomes
fully apparent when one attempts to repeat the analysis with a vanishing
cosmological constant, replacing the asymptotically \ads\ infinity
by an asymptotically flat infinity. We shall outline this analysis in
Appendix~\ref{app:as-flat}.  While there is no difficulty in
quantizing the reduced Hamiltonian theory, the trace of the analytically
continued time evolution operator turns out to remain divergent even
after a renormalization of the kind performed in Secs.\
\ref{subsec:grand-partition} and~\ref{sec:can-ensemble}\null. Neither
the canonical ensemble nor the grand canonical ensemble exists. For the
canonical ensemble this conclusion might be surprising in view of the
observation that a Reissner-Nordstr\"om black hole in asymptotically
flat space is stable against Hawking evaporation when one fixes the
charge and the temperature at the infinity, provided the mass and
charge parameters of the hole satisfy the inequality $\bq^2 >
\case{3}{4} \bm^2$ \cite{davies1}. However, as we shall see in
Appendix~\ref{app:as-flat}, the local stability of a classical solution
is not sufficient to guarantee the existence of a full thermodynamical
ensemble.

Finally, we recall that as the physical temperature of Hawking radiation
is redshifted to zero at the \ads\ infinity, we followed
Refs.\ \cite{pagerev,HPads,PagePhill} and defined a renormalized
temperature at infinity in terms of the rate at which the local Hawking
temperature approaches zero. This definition led to physically
reasonable conclusions; in particular, we recovered from the
thermodynamical ensembles the Bekenstein-Hawking result for the black
hole entropy. The definition can however be argued to have an {\em ad
hoc\/} flavor, and one might wish to replace it by something that can
be given a more immediate physical justification.\footnote{An
interesting possibility might be to {\em assume\/} the
Bekenstein-Hawking entropy and then to derive the appropriate
renormalized temperature \cite{BrownCreMann}.  However, it does not
appear clear how to adopt this as a starting point in a theory where the
Bekenstein-Hawking result is expected to be only an approximate one,
in the domain where the (grand) partition function is dominated by a
classical Euclidean solution.}
What would be needed is a better understanding as to whether
asymptotically \ads\ infinity can in some sense be regarded as a
physically realizable system, rather than just as a mathematically
elegant set of boundary conditions.

{\em Note added\/}.
After this work was completed, we became aware of Refs.\
\cite{kuns1,kuns2,kuns3}, which discuss the Dirac quantization of
four-dimensional spherically symmetric Einstein-Maxwell geometries and
related dilatonic theories. The work in these references has close
technical similarities to our work.

\acknowledgments
We would like to thank Dieter Brill, David Brown, John Friedman,
Domenico Giulini, Bei-Lok Hu, Werner Israel, Karel Kucha\v{r}, and
Kayll Lake for helpful discussions and correspondence. A~special thanks
is due to Bernard Whiting for valuable suggestions and a constructive
critique of the manuscript.  This work was supported in part by NSF
grants PHY91-05935, PHY91-19726, and PHY95-07740.

\appendix
\section{Reissner-Nordstr\"om-anti-de~Sitter black hole}
\label{app:rnads}

In this appendix we recall some relevant properties of the
\rnads\ (RNAdS) metric. We concentrate on the case where a nondegenerate
event horizon exists, and on the region exterior to this horizon.

In the curvature coordinates $(T,R)$, the RNAdS metric is
given by
\begin{mathletters}
\label{rnads-metric}
\begin{equation}
ds^2 = - F dT^2 + F^{-1} dR^2 + R^2 d\Omega^2
\ \ ,
\label{curv-metric}
\end{equation}
where $d\Omega^2$ is the metric on the unit two-sphere and
\begin{equation}
F := {R^2 \over \ell^2} + 1 - {2M \over R} + {Q^2 \over R^2}
\ \ .
\label{rnads-F}
\end{equation}
\end{mathletters}%
$T$ and $R$ are called respectively the Killing time and the
curvature radius. The parameter $\ell$ is positive, and we take the
parameters $M$ and $Q$ to be real. Together with the
electromagnetic potential
\begin{equation}
A = {Q \over R} dT
\ \ ,
\label{rnads-A}
\end{equation}
the metric~(\ref{rnads-metric}) is a solution to the Einstein-Maxwell
equations with the cosmological constant
$-3\ell^{-2}$ \cite{hoffmann,exact-book}. The parameters $M$ and $Q$
are referred to respectively as the mass and the (electric) charge.
The case $Q=0$ yields the Schwarzschild-anti-de~Sitter metric, and
the case $Q=M=0$ yields the metric on (the universal covering space of)
anti-de~Sitter space \cite{haw-ell}.

The metric (\ref{rnads-metric}) has an asymptotically \ads\ infinity
at $R\to\infty$ for all values of the parameters \cite{HennTeit-ads}.
We wish to restrict the parameters so that the metric describes the
exterior of a black hole with a nondegenerate horizon. This happens when
the quartic polynomial $R^2 F(R)$ has a simple positive root $R=R_0$,
such that $F$ is positive for $R>R_0$. The necessary and sufficient
condition is $M>M_{\rm crit}(Q)$, where
\begin{equation}
M_{\rm crit}(Q) := {\ell \over 3 \sqrt{6} }
\left( \sqrt{1 + 12 {(Q/\ell)}^2} +2 \right)
{\left( \sqrt{1 + 12 {(Q/\ell)}^2} -1 \right)}^{1/2}
\ \ .
\label{McritQ}
\end{equation}
Note that $M$ is then necessarily positive. $R_0$~can now be
determined uniquely as the function $\Rhor(M,Q)$ of $M$
and~$Q$: for $Q=0$, $\Rhor(M,Q)$ is defined as the unique positive
solution to the equation $F=0$; for $Q\ne0$, $\Rhor(M,Q)$ is defined
as the larger of the two positive solutions. In either case, if $Q$ is
considered fixed, $\Rhor(M,Q)$ is a monotonically increasing
function of $M$ that takes the values $\Rcrit(Q) < \Rhor(M,Q) <
\infty$ as $M_{\rm crit}(Q) < M < \infty$, where
\begin{equation}
\Rcrit(Q) :=
{\ell\over\sqrt{6}}
\left( \sqrt{1 + 12 {(Q/\ell)}^2} -1 \right)^{1/2}
\ \ .
\label{Rcrit}
\end{equation}

The metric can thus be uniquely parametrized by $Q$
and~$R_0$. The only restriction for these parameters is
\begin{equation}
R_0 > \Rcrit(Q)
\ \ ,
\end{equation}
and the mass is then given by
\begin{equation}
M = {R_0 \over 2}
\left( {R_0^2 \over \ell^2} + 1 + {Q^2 \over R_0^2} \right)
\ \ .
\label{massfunc}
\end{equation}

With $R_0 < R < \infty$, the metric (\ref{curv-metric}) covers the
region from the horizon to the asymptotically \ads\ infinity. The
Penrose diagram can be found in Refs.\ \cite{btz-cont,lake1}.

\section{Reissner-Nordstr\"om black hole in asymptotically flat
space}
\label{app:as-flat}

In the main text we took the cosmological constant to be strictly
negative. In this appendix we shall outline the corresponding classical
and quantum mechanical analysis in the case where the cosmological
constant vanishes. In the notation of the main text this means taking
the limit $\ell\to\infty$. The classical solutions are then not
asymptotically \ads\ but asymptotically flat, and the falloff
conditions at $r\to\infty$ must be modified to reflect this fact.

In the variables of Section~\ref{sec:metric}, we retain the
falloff conditions (\ref{s-r}) at $r\to0$, but at $r\to\infty$ we
introduce the new falloff conditions
\begin{mathletters}
\label{flat:l-r}
\begin{eqnarray}
\Lambda(t,r) &=&
1 + M_+(t) r^{-1} +
O^\infty (r^{-1-\epsilon})
\ \ ,
\label{flat:l-r-Lambda}
\\
R(t,r) &=& r + O^\infty (r^{-\epsilon})
\ \ ,
\label{flat:l-r-R}
\\
P_{\Lambda}(t,r) &=& O^\infty (r^{-\epsilon})
\ \ ,
\label{flat:l-r-PLambda}
\\
P_{R}(t,r) &=& O^\infty (r^{-1-\epsilon})
\ \ ,
\label{flat:l-r-PR}
\\
N(t,r) &=& N_+(t) + O^\infty(r^{-\epsilon})
\ \ ,
\label{flat:l-r-N}
\\
N^r(t,r) &=& O^\infty (r^{-\epsilon})
\ \ ,
\label{flat:l-r-Nr}
\\
\Gamma(t,r) &=& O^\infty (r^{-1-\epsilon})
\ \ ,
\label{flat:l-r-B}
\\
P_\Gamma(t,r) &=& Q_+(t) + O^\infty (r^{-\epsilon})
\ \ ,
\label{flat:l-r-PB}
\\
{\tilde \Phi}(t,r) &=& {\tilde \Phi}_+(t) + O^\infty (r^{-\epsilon})
\ \ ,
\label{flat:l-r-tPhi}
\end{eqnarray}
\end{mathletters}%
where $0<\epsilon\le1$. For the metric quantities these conditions are
precisely those used in Ref.\ \cite{kuchar1}, ensuring asymptotic
flatness. These conditions make the bulk action (\ref{S-ham})
well-defined, and  they are preserved under the time evolution. Adding
the boundary action
\begin{equation}
\int dt \left(
\casehalf R_0^2 N_1 \Lambda_0^{-1} - N_+ M_+ + {\tilde \Phi}_0
Q_0 - {\tilde \Phi}_+ Q_+ \right)
\label{flat:b-term}
\end{equation}
yields an action for a variational principle in which $N_+$,
$N_1\Lambda_0^{-1}$, ${\tilde \Phi}_+$, and ${\tilde \Phi}_0$ are
prescribed functions of~$t$. Dropping the last two terms in
(\ref{flat:b-term}) yields an action for a variational principle in
which ${\tilde \Phi}_+$ and ${\tilde \Phi}_0$ are free but $Q_+$ and
$Q_0$ are prescribed.

The canonical transformation of the main text can now be adapted to the
present boundary conditions by simply taking the limit $\ell\to\infty$.
A~new action can be constructed as in Sec.~\ref{subsec:action},
the new falloff conditions only giving rise to minor technical
modifications to the redefinition of the Lagrange multipliers. In the
theory that prescribes ${\tilde\Phi}_+$ and~${\tilde\Phi}_0$,
elimination of the constraints along the lines of Sec.\
\ref{sec:reduction} yields the reduced action
\begin{equation}
S [ \bm, \bq, {\bf p}_\bm, {\bf p}_\bq ;
{\tilde N}^M_0 , N_+ , {\tilde \Phi}_+ , {\tilde \Phi}_0 ]
=
\int dt
\left(
{\bf p}_\bm {\dot \bm}
+
{\bf p}_\bq {\dot \bq}
- {\bf h} \right)
\ \ ,
\label{flat:S-red}
\end{equation}
where the reduced Hamiltonian is given by
\begin{equation}
{\bf h} = - \casehalf \Rh^2 {\tilde N}^M_0
+ N_+ \bm +
\left( {\tilde \Phi}_+ - {\tilde \Phi}_0 \right) \bq
\label{flat:h-red}
\end{equation}
with $\Rh := \bm + \sqrt{\bm^2 - \bq^2}$. The range of the variables is
$0<\bm$, $\bq^2 < \bm^2$. In the theory that prescribes $Q_+$
and~$Q_0$, one proceeds as in Sec.\ \ref{sec:can-ensemble} to obtain
the reduced action
\begin{equation}
S_C [ \bm, {\bf p}_\bm ; {\tilde N}^M_0 , N_+ ; \bq ]  =
\int dt
\left( {\bf p}_\bm {\dot \bm} - {\bf h}_C \right)
\ \ ,
\label{flat:SC-red}
\end{equation}
where $\bq$ is now regarded as an external parameter and
\begin{equation}
{\bf h}_C = - \casehalf \Rh^2 {\tilde N}^M_0  +
N_+ \bm
\ \ .
\label{flat:hC-red}
\end{equation}

Quantization of the two reduced theories proceeds as in the main text.
For the renormalized trace of the analytically continued time
evolution operator, we obtain formally
\begin{mathletters}
\label{flat:partitions}
\begin{eqnarray}
\czren (\beta,\phi)
&=& {\cal N} \int\limits_{\Rh>|\bq|} {\tilde \mu}
\, d\Rh d\bq \,
\exp\left(  -I_* \right)
\ \ ,
\label{flat:cZ-istar}
\\
\zren (\beta,\bq)
&=& \int\limits_{|\bq|}^\infty
{\tilde{\tilde\mu}}
\, d\Rh \,
\exp\left(  -I_{C*} \right)
\ \ ,
\label{flat:Z-istarc}
\end{eqnarray}
\end{mathletters}%
where $I_*$ and $I_{C*}$ are respectively given by dropping the term
proportional to $\ell^{-2}$ from Eqs.\ (\ref{istar}) and~(\ref{istarc}).
$\beta$~is now interpreted as the inverse Hawking temperature at the
infinity, with no renormalization. However, both integrals in
(\ref{flat:partitions}) are divergent because of the behavior of $I_*$
and $I_{C*}$ at large~$\Rh$. Thus, neither the canonical ensemble nor
the grand canonical ensemble exists under the asymptotically flat
boundary conditions. In this respect, the inclusion of the charge has
therefore not made a qualitative difference from the asymptotically
flat vacuum case \cite{LW2}.

The critical points of $I_*$ and $I_{C*}$ give again the (Lorentzian)
classical solutions that have the inverse Hawking temperature
$\beta$ at infinity and the prescribed value of respectively
$\phi$ or~$\bq$. The condition that $I_*$ possess critical points is
$|\phi|<1$: when this condition is satisfied there exists
exactly one critical point, but this critical point is not a local
extremum. This reproduces the observations made by Davies in
Ref.\ \cite{davies1,davies2} about charged black hole equilibria with
fixed~$\phi$, and reflects in particular the fact that a semiclassical
charged black hole under these boundary conditions is not stable
against Hawking evaporation.

The condition that $I_{C*}$ possess critical points is $\beta/|\bq|
\ge 6\pi\sqrt{3}$, and when the inequality is genuine, there exist two
critical points. The critical point with the smaller (larger) value of
$\Rh$ is a local minimum (maximum, respectively). The local minimum
satisfies $\bq^2 > \case{3}{4} \bm^2$, and it corresponds to the
classical solution that Davies \cite{davies1} showed to be stable
against Hawking evaporation under these boundary conditions (see also
Refs.\ \cite{lousto1,lousto2,lousto3}). While the
thermodynamical stability of this semiclassical solution is reflected
in its being a local minimum of our~$I_{C*}$ \cite{whitingCQG}, the
divergence of the the integral in (\ref{flat:Z-istarc}) demonstrates
that this local stability is not sufficient to guarantee the existence
of a full thermodynamical canonical ensemble.

\end{document}